\documentclass[lettersize,journal]{IEEEtran}
\usepackage{amsmath,amsfonts}
\usepackage{algorithmic}
\usepackage{algorithm}
\usepackage{array}
\usepackage[caption=false,font=normalsize,labelfont=sf,textfont=sf]{subfig}
\usepackage{textcomp}
\usepackage{stfloats}
\usepackage{url}
\usepackage{verbatim}
\usepackage{graphicx}
\usepackage{cite}
\usepackage{epstopdf}
\usepackage{caption2}
\usepackage{subfig}
\usepackage{amssymb}
\usepackage{makecell}
\usepackage{xcolor}
\usepackage{bbding}
\usepackage[colorlinks,
            linkcolor=blue,
            citecolor=blue,
            urlcolor=black,
            anchorcolor=black,
            pdfborder={0 0 0}]{hyperref}
\begin{document}

\title{Division-based Receiver-agnostic RFF Identification in WiFi Systems}

\author{\IEEEauthorblockN{Xuan Yang,~\IEEEmembership{Graduate Student Member, IEEE}, Dongming Li,~\IEEEmembership{Member, IEEE}, Dong Wei, Meng Zhang}

\thanks{This work has been submitted to the IEEE for possible publication. Copyright may be transferred without notice, after which this version may no longer be accessible. This work was partially supported by the Natural Science Funding of Jiangsu Province under Grant No. BK20252011, and the Big Data Computing Center of Southeast University \emph{(Corresponding author: Dongming Li.)}}

\thanks{X. Yang and D. Li are with School of Cyber Science and Engineering, Southeast University, Nanjing, 211189, China. (email: xuan\_yang@seu.edu.cn; lidm@seu.edu.cn).}

\thanks{D. Wei and M. Zhang are with the Institute of Information Engineering, Chinese Academy of Sciences, Beijing 100085, China (e-mail: weidong@iie.ac.cn; zhangmeng@iie.ac.cn).}%
}

\maketitle

\begin{abstract}
In physical-layer security schemes, radio frequency fingerprint (RFF) identification of WiFi devices is susceptible to receiver differences, which can significantly degrade classification performance when a model is trained on one receiver but tested on another. Consequently, the receiver-agnostic physical-layer identification of WiFi devices remains an important research challenge. Recent studies have addressed this issue by either training on signals collected from multiple receivers to extract receiver-agnostic RFF features, or by employing calibration processes to enhance classification accuracy. However, these approaches introduce substantial computational overhead. In this paper, we propose a division-based receiver-agnostic RFF extraction method for WiFi systems, which removes the receivers' effects by dividing different preambles in the frequency domain. The proposed method requires only a single receiver for training and does not rely on additional calibration or stacking processes. First, for flat fading channel scenarios, the legacy short training field (L-STF) and legacy long training field (L-LTF) of the unknown device are divided by those of the reference device in the frequency domain. The receiver-dependent effects can be eliminated with the requirement of only a single receiver for training, and the higher-dimensional RFF features can be extracted. Second, for frequency-selective fading channel scenarios, the high-throughput long training field (HT-LTF) is divided by the L-LTF in the frequency domain. Only a single receiver is required for training and the higher-dimensional RFF features that are both channel-invariant and receiver-agnostic are extracted. Finally, simulation and experimental results demonstrate that the proposed method effectively mitigate the impacts of channel variations and receiver differences. The classification results show that, even when training on a single receiver and testing on a different one, the proposed method achieves classification accuracy improvements of 15.5\% and 28.45\% over the state-of-the-art approach in flat fading and frequency-selective fading channel scenarios, respectively.
\end{abstract}


\vspace{-0.3cm}
\section{Introduction}
\IEEEPARstart{O}{ver} the past several decades, the rapid development of WiFi devices has changed various aspects of daily life. As the number of WiFi devices continues growing, ensuring secure WiFi communications has become increasingly important \cite{babun2021survey}. Traditionally, wireless system security is achieved through encryption algorithms and security protocols. However, with the rise of computational power and the advent of quantum computing, security architectures that are based on computational complexity face growing threats \cite{stoyanova2020survey}, because such systems can be easily cracked by quantum computing.

Radio frequency fingerprint (RFF) identification is a promising technology for device identification, which does not rely on computational complexity and can enhance authentication security. The transmitted signal of a device is inherently affected by various hardware imperfections, such as in-phase/quadrature (IQ) direct current (DC) offset, IQ imbalance, imperfect low-pass filtering, and power amplifier (PA) nonlinearity, all of which result in a distorted signal \cite{yang2023led}. These unique hardware impairments, collectively referred to as RFF, can be extracted from the distorted signal. Similar to a human fingerprint, the RFF is extractable, unique, and invariant, making it a reliable feature for achieving authentication security \cite{abbas2024radio}.

Several key challenges must be addressed to enable the practical deployment of RFF identification systems. First, wireless channel effects can distort the transmitted signal. When the training and test sets are collected from different locations or at different times, classification accuracy can degrade substantially. To mitigate this issue, researchers have proposed methods to extract inherent RFF features \cite{hou2014physical, vo2016fingerprinting, hua2018accurate, liu2019real, kandel2019exploiting}, which are independent of channel effects. Additionally, channel estimation and compensation techniques \cite{jian2020deep, jian2021radio}, as well as filtering based approaches \cite{chen2022radio, chen2023channel, yang2023led, kong2024csi}, have been developed to directly counteract channel effects. However, channel estimation and compensation may inevitably alter the original RFF, while the filtering based approaches face difficulties in accurately determining the passband and stopband. Moreover, by leveraging the properties of channel coherence time \cite{chen2022radioDSRC, xing2022design, yang2023use, zhang2024channel} and coherence bandwidth \cite{he2023radio, he2023channel, he2024radio}, signal processing techniques based on the convolution model can be applied to mitigate channel distortions. Furthermore, data augmentation methods \cite{soltani2020more, al2021deeplora, xing2022design, comert2022analysis} can enhance classification accuracy, although they typically incur increased training overhead.

Second, the transmitted signal is also influenced by the receiver's RFF, leading to degraded classification accuracy when the receiver changes. Several neural network based methods \cite{chen2021radio, gaskin2022tweak, zheng2022method, huang2022cross, gaskin2023deep, zha2023cross, zhao2023gan, shen2023towards, bao2023receiver, chen2024cross, yu2024rf} have been proposed to extract receiver-agnostic RFF features. However, training a receiver-agnostic RFF feature extractor typically requires either collecting training samples from multiple receivers or performing a calibration process when deploying on a new receiver. Both approaches are time-consuming. To address this issue, the authors in \cite{xing2022design, chen2022radioDSRC} proposed a division-based RFF extraction method that separates the short and long preambles in the frequency domain to obtain receiver-agnostic features. However, this method can extract only 12-dimensional RFF features, as the short preamble in WiFi and dedicated short-range communication (DSRC) frames occupies just 12 subcarriers, which limits classification performance. Additionally, the method in \cite{xing2022design, chen2022radioDSRC} relies on a stacking process \cite{xing2018radio}, where adjacent frames are combined for denoising and improved classification accuracy. As adjacent frames may originate from different devices, the stacking process is difficult to implement reliably in practice. Therefore, achieving high classification accuracy with training on only a single receiver while eliminating the requirements of stacking and calibration processes deserves further investigation.

Although the afore-mentioned research progress, the existing studies on receiver-agnostic RFF identification have the following drawbacks: (1) The training process typically requires multiple receivers, making both data collection and classifier training time-consuming. (2) A calibration process is needed when deploying on a new receiver, introducing additional overhead. (3) Only low dimensional RFF features are extracted, leading to reduced classification accuracy. (4) A stacking process is often required, which causes delays in delivering the final identification results. To address these limitations and mitigate the adverse receiver effects, we model the adverse receiver effects and propose a division-based receiver-agnostic RFF identification method. To the best of our knowledge, this is the first work that eliminates the need for multiple receivers, the calibration process, and the stacking process, while enabling the extraction of high dimensional RFF features to improve classification accuracy. The main contributions of this paper are summarized as follows:

1. For flat fading channel scenarios, the legacy short training field (L-STF) and legacy long training field (L-LTF) of the unknown device are divided by those of the reference device in the frequency domain. The receiver-dependent effects can be eliminated with the requirement of only a single receiver for training, and the higher-dimensional RFF features can be extracted.

2. For frequency-selective fading channel scenarios, the aforementioned reference device based approach fails to fully mitigate channel effects. To address this issue, the high-throughput long training field (HT-LTF) is divided by the L-LTF in the frequency domain. This approach operates without a reference device, requires only a single receiver for training, extracts higher-dimensional RFF features that are both channel-invariant and receiver-agnostic, and achieves high classification accuracy.

3. Extensive experiments are conducted on commercial WiFi devices, demonstrating that by dividing different preambles in the frequency domain, the adverse receiver effects can be removed, without requiring multiple receivers, stacking, or calibration processes. Specifically, the proposed method achieves classification accuracies of up to 98.47\% in the flat fading channel scenario and 94.91\% in the frequency-selective fading channel scenario, even when the receiver changes.

Similar to existing works such as \cite{chen2022radioDSRC, xing2022design, shen2023toward, shen2023towards, xie2025towards}, this paper primarily focuses on closed-set classification, where all test devices are known during the training stage. For open-set classification, in which rogue devices may appear during the testing stage, readers are recommended to refer to \cite{xie2021generalizable, li2025meta}. The corresponding dataset \footnote{\href{https://ieee-dataport.org/documents/wifirffi}{https://ieee-dataport.org/documents/wifirffi}} of this work is available online.

The remainder of this paper is structured as follows: Section \ref{section:2} provides an overview of the related work in this field. Section \ref{section:3} elaborates on WiFi signal preprocessing. Section \ref{section:4} presents the proposed receiver-agnostic RFF feature extraction and identification methods. Section \ref{section:5} evaluates the proposed methods through experimental analysis. Finally, Section \ref{section:6} concludes the paper.

\section{Related Work}
\label{section:2}
Since channel variations and receiver differences are the two main challenges in RFF identification, this section primarily focuses on addressing these two issues.

\subsection{Channel Robust RFF identification}
Since wireless signals propagate through open-air environments, removing channel effects is essential to obtain environment-independent RFF features. The inherent RFF, such as carrier frequency offset (CFO) \cite{hou2014physical, vo2016fingerprinting, hua2018accurate}, is assumed to be channel-independent and can be extracted for classification purposes. However, long-term experiments have shown that the CFO characteristic is unstable over time, and the CFO characteristic is also prone to be affected by the temperature of the crystal oscillator \cite{gu2022terff}. Channel estimation and compensation methods \cite{jian2020deep, jian2021radio} have been employed to mitigate channel effects and derive channel-robust RFF. However, the channel estimation and compensation process may unintentionally impair the RFF features, leading to lower classification accuracy. Under the assumption that multipath signals have higher energy in the taps with smaller delay, some filtering based methods are designed to extract RFF features \cite{chen2022radio, chen2023channel, yang2023led, kong2024csi}. Additionally, division-based RFF extraction methods, exploiting the characteristics of channel coherent time \cite{chen2022radioDSRC, xing2022design, yang2023use, zhang2024channel} and channel coherent bandwidth \cite{he2023radio, he2023channel, he2024radio}, have been proposed to extract channel-independent RFF. Moreover, the authors in \cite{restuccia2019deepradioid, restuccia2021deepfir} compensated channel effects by optimizing the digital finite impulse response (FIR) filter on the transmitter side. Furthermore, data augmentation methods can be utilized to achieve higher classification accuracy \cite{soltani2020more, al2021deeplora, xing2022design, comert2022analysis}, though they lead to significantly increased training overhead.
\begin{figure*}[htbp]
    \centering
    \includegraphics[width=12cm]{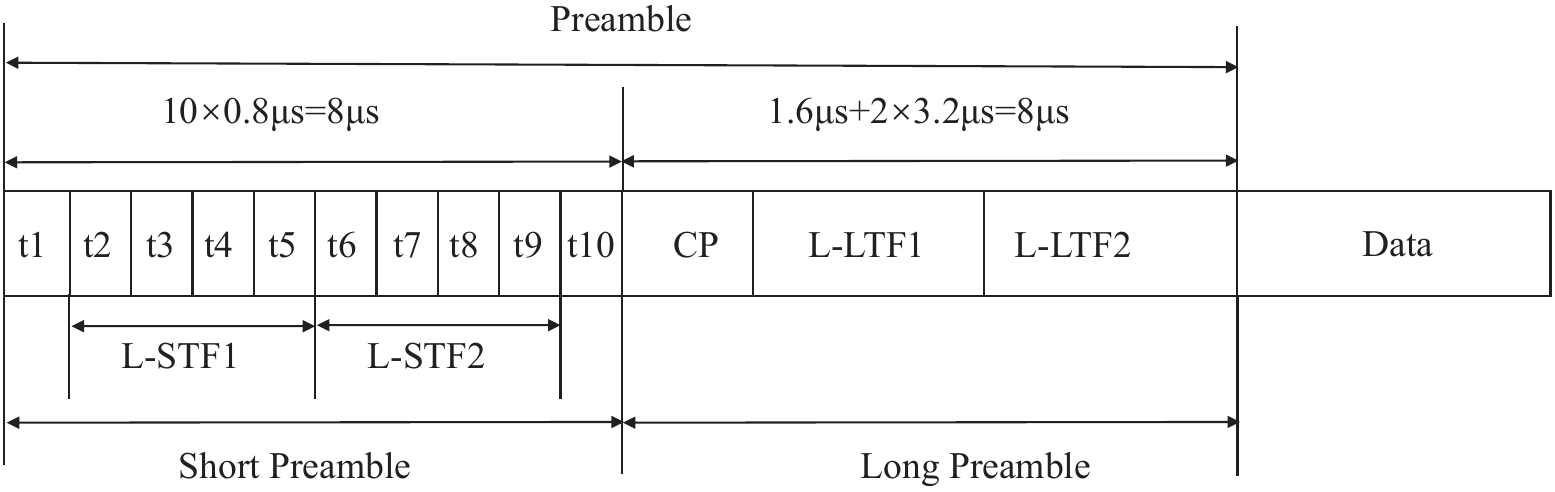}
    \vspace{-0.2cm}
    \caption{Non-HT PPDU structure.}
    \label{fig1}
    \vspace{-0.1cm}
\end{figure*}

\begin{figure*}[htbp]
    \centering
    \includegraphics[width=13cm]{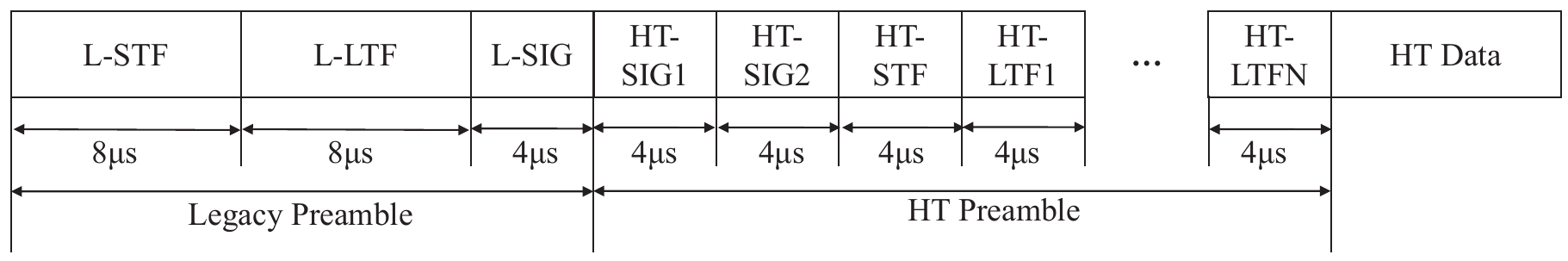}
    \vspace{-0.2cm}
    \caption{HT-MF PPDU structure.}
    \label{fig2}
    \vspace{-0.6cm}
\end{figure*}

\subsection{Receiver-agnostic RFF Identification}
After the transmitted signal is received by the receiver, the transmitter's RFF features are also influenced by the receiver's characteristics. Since receiver differences decrease classification accuracy \cite{elmaghbub2021lora, hanna2022wisig}, handling receiver impacts is crucial, and has been explored in several studies. Various neural network based methods, such as convolutional neural networks (CNN) \cite{zheng2022method, chen2024cross}, Siamese networks \cite{gaskin2023deep}, simple Siamese \cite{zha2023cross}, generative adversarial networks (GAN) \cite{zhao2023gan, yu2024rf}, adversarial training \cite{shen2023towards}, deep adversarial neural networks (DANN) \cite{huang2022cross, bao2023receiver}, transfer learning \cite{chen2021radio}, and deep metric learning \cite{gaskin2022tweak}, have been utilized to tackle this problem. To enable the neural network to extract receiver-agnostic RFF features, many studies require signals received from multiple receivers \cite{bao2023receiver, zha2023cross, liu2023receiver, shen2023towards, zhao2023gan}, which increases the overhead in the signal collection process. Furthermore, for domain adaptation and enhancing classification accuracy, a calibration process involving labeled or unlabeled data from the new receiver is required in \cite{merchant2019toward, chen2021radio, gaskin2022tweak, huang2022cross, zheng2022method, bao2023receiver, zha2023cross, gaskin2023deep, shen2023towards, chen2024cross, yang2024mitigating, yu2024rf}. Since the neural network needs to be recalibrated for each new receiver, calibration-based methods are time consuming. Alternatively, instead of using neural networks, the authors in \cite{abdollahi2024novel} proposed a joint channel and RFF estimation method that accounts for receiver effects by exploiting wave propagation theory and a compressive sensing linear solution. However, this approach requires receiver calibration and prior knowledge of the transmitter's precise location, limiting its universality.

The division-based RFF extraction methods \cite{chen2022radioDSRC, xing2022design, zhang2024channel}, which do not require training data collected from multiple receivers or a calibration process, have been theoretically proven to mitigate the adverse receiver effects \cite{chen2022radioDSRC}. However, since the adjacent L-LTF parts are similar, the features extracted by dividing these two parts are susceptible to noise effects. These adverse effects are also observed in \cite{zhang2024channel}, where a multiple discriminant analysis/maximum likelihood (MDA/ML) classifier is employed for classification. By dividing the short preamble from the long preamble in the frequency domain, the channel and receiver impacts can be eliminated \cite{xing2022design, chen2022radioDSRC}. Nevertheless, since the short preambles in WiFi \cite{xing2022design} and DSRC \cite{chen2022radioDSRC} systems occupy only 12 subcarriers, the extracted RFF features are limited to just 12 dimensions, which leads to lower classification accuracy. Moreover, as the RFF features of the short preamble and long preamble are similar, the extracted features in \cite{xing2022design, chen2022radioDSRC} are unstable and vulnerable to noise, with the stacking process being required to improve accuracy. However, the stacking process is challenging to implement since adjacent frames may originate from different devices.
\begin{table}[h!t]
\vspace{-0.1cm}
\scriptsize
   \centering
   \caption{Related Work}
    \begin{center}
     \begin{tabular}{||c c c c||}
     \hline
     Paper & {\makecell[c] {Need Multiple \\ Receivers?}} & {\makecell[c] {Need Calibration \\ Process?}} & Method\\
     \hline
     \cite{bao2023receiver} & Yes & Yes & DANN\\
     \hline
     \cite{zha2023cross} & Yes & Yes & SimSiam\\
     \hline
     \cite{shen2023towards} & Yes & Yes & Adversarial Training\\
     \hline
     \cite{liu2023receiver} & Yes & No & Loss Function Design\\
     \hline
     \cite{zhao2023gan} & Yes & No & GAN \\
     \hline
     \cite{merchant2019toward} & No & Yes & Transform Network\\
     \hline
     \cite{chen2021radio} & No & Yes & Transfer Learning\\
     \hline
     \cite{gaskin2022tweak} & No & Yes & Deep Metric Learning\\
     \hline
     \cite{huang2022cross} & No & Yes & DANN\\
     \hline
     \cite{zheng2022method} & No & Yes & CNN\\
     \hline
     \cite{gaskin2023deep} & No & Yes & Siamese\\
     \hline
     \cite{chen2024cross} & No & Yes & CNN\\
     \hline
     \cite{yang2024mitigating} & No & Yes & Adaptive Pseudo-labeling\\
     \hline
     \cite{yu2024rf} & No & Yes & Dual Transfer GAN\\
     \hline
     \cite{abdollahi2024novel} & No & Yes & Compressive Sensing\\
     \hline
     \cite{chen2022radioDSRC} & No & No & Division, Random Forest\\
     \hline
     \cite{xing2022design} & No & No & Division, CNN\\
     \hline
     \cite{zhang2024channel} & No & No & Division, MDA/ML\\
     \hline
     Ours & No & No & Division, InceptionTime\\
     \hline
    \end{tabular}
    \end{center}
\label{tab1}
\vspace{-0.5cm}
\end{table}
Furthermore, while the effectiveness of the receiver-agnostic RFF extraction method in \cite{chen2022radioDSRC} has been demonstrated theoretically, its performance in real-world scenarios deserves verification.

The related works in the receiver-agnostic research field are summarized in Table \ref{tab1}. It can be found that most of the existing works need either multiple receivers or the calibration process, or both simultaneously. Only a few works \cite{chen2022radioDSRC, xing2022design, zhang2024channel} require neither multiple receivers nor calibration process. Most of these existing works \cite{xing2022design, zhang2024channel} are designed to remove channel effects and mitigate the receiver's effects inadvertently. Only one of these existing works \cite{chen2022radioDSRC} theoretically demonstrated a division-based receiver-agnostic RFF extraction method, the real-world performance is under verification.
\section{WiFi Frame Structure and Signal Preprocessing}
\label{section:3}
To obtain receiver-agnostic RFF features, signal preprocessing is essential. In this section, WiFi frame structure and signal preprocessing are introduced in detail, while the signal preprocessing involves signal detection, frame synchronization, and CFO compensation.
\subsection{WiFi Frame Structure}
The physical layer protocol data unit (PPDU) formats of non-high-throughput (non-HT) and high-throughput mixed format (HT-MF), shown in Fig. \ref{fig1} and Fig. \ref{fig2}, are primarily utilized for RFF extraction in this work. In Fig. \ref{fig1}, the non-HT frame consists of a preamble and a data portion. The WiFi preamble is mainly used for signal synchronization and channel estimation. The preamble, which is invariant and can be used for RFF extraction, includes ten repeated short training symbols, i.e., the legacy short training field (L-STF) part, a cyclic prefix (CP), and two repeated long training symbols, i.e., the L-LTF part. The L-STF part occupies 12 subcarriers and lasts 8 $\mu$s in the time domain, corresponding to 160 samples at a sampling rate of 20 Msps. The L-LTF part occupies 52 subcarriers, and the CP lasts 1.6 $\mu$s, while each of the replicated symbols lasts 3.2 $\mu$s. Thus, at a sampling rate of 20 Msps, the CP part contains 32 samples, and each long training symbol contains 64 samples. The signal duration used for fast Fourier transform (FFT) is 3.2 $\mu$s or 64 samples at the 20 Msps sampling rate. We define the 17th to the 80th samples and the 81st to the 144th samples of the L-STF part as L-STF1 and L-STF2, respectively. Similarly, the 33rd to the 96th samples and the 97th to the 160th samples of the L-LTF part are denoted as L-LTF1 and L-LTF2, respectively. Since L-STF1 and L-STF2 are identical, and L-LTF1 and L-LTF2 are also identical, L-STF1 and L-STF2 can be combined to reduce the adverse effects of noise. The same operation is applied to L-LTF1 and L-LTF2.

In Fig. \ref{fig2}, the HT-MF frame consists of a legacy preamble, an HT preamble, and a data portion. The legacy preamble of the HT-MF frame is identical to that of the non-HT frame. In the HT preamble, the number of HT-LTF may vary. Each HT-LTF occupies 54 subcarriers and contains 80 samples at a sampling rate of 20 Msps. Since the HT-LTF part is invariant, it can be used for RFF extraction. To avoid inter-symbol interference, the 17th to the 80th samples of the HT-LTF part, denoted as HT-LTF1, are used for RFF extraction.

\subsection{Signal Detection}
In this procedure, the signal's energy is compared with a threshold $T$, which should be set according to real world conditions. The equivalent baseband signal can be denoted as $y\left( n \right)$, where $n = 1,2,...,N$, with $N$ representing the length of samples. The number of samples utilized for the signal energy calculation is denoted as $W$. In the ${k}$-th detection round, the coarse signal starting point is detected when the following condition is met
\begin{equation}
\label{eq1}
E = \sum\limits_{w = 1}^W {\left| {y\left( {\left( {k - 1} \right)W + w} \right)} \right|}  > T.
\end{equation}
It can be obtained that the starting point is approximate ${n_0} = \left( {k - 1} \right)W$, since the beginning of the signal causes the value of $E$ to exceed the threshold $T$.

\subsection{Frame Synchronization}
The correlation-based method is used for frame synchronization. The ideal signal of the L-LTF part is generated offline and can be represented as $L\left( n \right)$, $n = 1,...,{L_L}$, with ${L_L}$ denoting the length of the L-LTF part. The starting point of the L-LTF part can be obtained as
\begin{equation}
\label{eq2}
{k_0} = \arg \mathop {\max }\limits_k \sum\limits_{n = k}^{k + {L_L} - 1} {y\left( {{n_0} + n} \right)} {L^*}\left( {n - k + 1} \right),k = 1,...,K,
\end{equation}
where ${\left(  \cdot  \right)^*}$ denotes the conjugate operation and $K$ is the search length. Then, the starting point of the frame can be obtained according to the frame structure as ${n_1} = {k_0} - 160$ under a sampling rate of 20 Msps, since there are 160 samples between the starting point of the frame and the starting point of the L-LTF part.

\begin{figure*}[htbp]
    \centering
    \subfloat{
        \includegraphics[width=13cm]{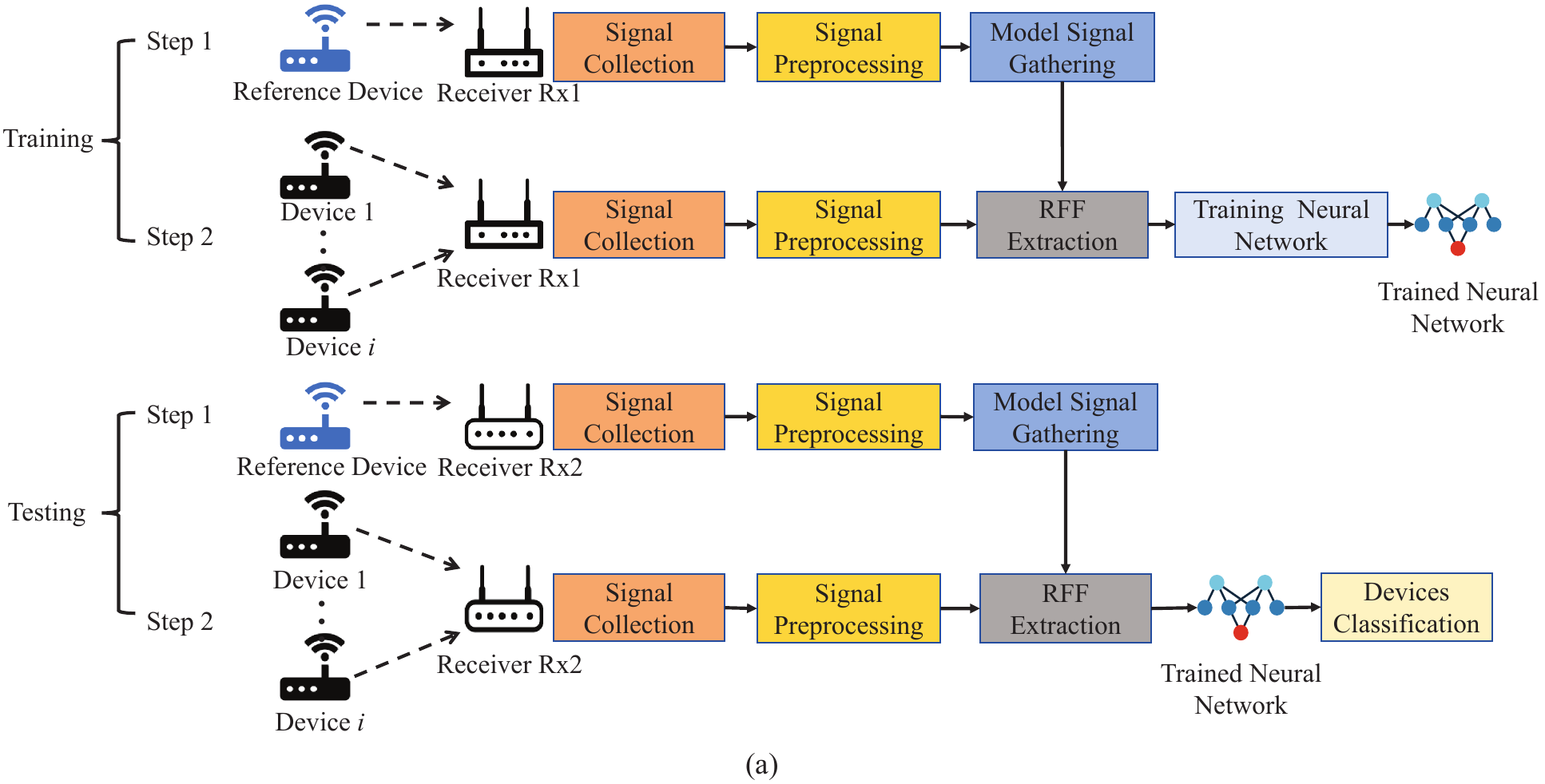}
        }\hfill
    \subfloat{
        \includegraphics[width=13cm]{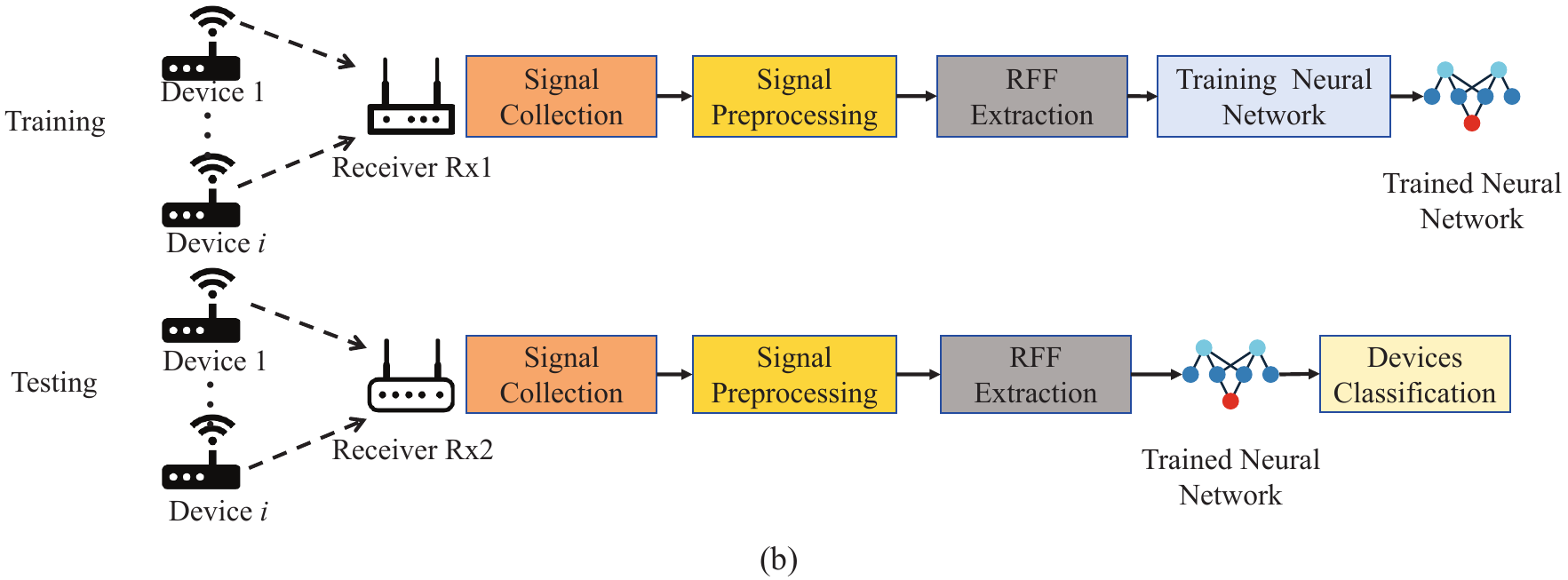}
        }
    \vspace{-0.1cm}
    \caption{Diagrams illustrating the proposed receiver-agnostic RFF extraction and classification method: (a) Under flat fading channel scenarios and (b) under frequency-selective fading channel scenarios.}
    \label{fig3}
    \vspace{-0.6cm}
\end{figure*}

\subsection{CFO Compensation}
The discrepancy between the transmitter's and receiver's central carrier frequencies results in the CFO. Ignoring channel effects and additive noise, the signal affected by the CFO can be expressed as
\begin{equation}
\label{eq3}
y(n) = x\left( n \right){e^{j2\pi \left( {{f_1} - {f_2}} \right)t}} = x(n){e^{j2\pi \Delta fn{T_s}}},
\end{equation}
where $x\left( n \right)$ is the transmitted baseband signal, and ${f_1}$ and ${f_2}$ are the carrier frequencies of the transmitter and receiver, respectively. The CFO can be estimated using the duplicate symbols of the L-STF part as
\begin{equation}
\label{eq4}
\hat f = \frac{1}{{2\pi {T_S}D}}\angle \sum\limits_{n = {n_S}}^{8D + {n_S} - 1} {{y^*}\left( {{n_1} + n} \right)y\left( {{n_1} + n + D} \right)},
\end{equation}
where $D$ is the symbol length of L-STF, and ${n_S}$ is the starting point used for the CFO estimation process, where $0 \le {n_S} \le D$. $T_S$ is the sampling rate, and $\angle  \cdot $ represents taking the phase of a complex number. There are totally eight repeated short training symbols utilized for CFO estimation. The signal after CFO compensation can be expressed as
\begin{equation}
\label{eq5}
{y_1}\left( n \right) = y\left( n \right){e^{ - j2\pi \widehat fn{T_S}}} \approx x\left( n \right).
\end{equation}
To compensate for the CFO more effectively, a fine CFO compensation method can be adopted using the duplicated structure of the L-LTF. This process is similar to the CFO compensation process with the L-STF part, and the detailed description is omitted for brevity.

\section{RFF Extraction and Identification}
\label{section:4}
The RFF extraction method for flat fading channel and frequency-selective fading channel scenarios is proposed. For flat fading channel scenarios, a reference device based, receiver-agnostic RFF identification method, referred to as the RD method, is designed to mitigate the receiver's RFF effects. For frequency-selective fading channel scenarios, a division-based receiver-agnostic RFF identification method, referred to as the HL method, is proposed to remove both the channel and the receiver's RFF effects, without the need for a reference device.
\subsection{System Overview}
The system diagrams of the proposed method are shown in Fig. \ref{fig3}. Fig. \ref{fig3}(a) illustrates the RFF extraction process for flat fading channel scenarios. During the training stage, receiver Rx1 first receives the signal from the reference device, referred to as the model signal. Then, Rx1 collects signals from various devices and extracts receiver-agnostic RFF features with the aid of the model signal. These extracted RFF features are then used to train a neural network. In the test stage, a new receiver, Rx2, first gathers the model signal from the reference device. Next, Rx2 collects signals from different devices and extracts receiver-agnostic RFF features with the help of the model signal. These extracted RFF features are then input into the trained neural network for classification. The RFF extraction process for frequency-selective fading channel scenarios is shown in Fig. \ref{fig3}(b). In the training stage, Rx1 collects signals from various devices and directly extracts the receiver-agnostic RFF features to train a neural network. During the test stage, Rx2 receives signals from different devices, extracts the receiver-agnostic RFF features, and uses the trained neural network for classification.
\subsection{RFF Extraction under Flat Fading Channel Scenarios}
\subsubsection{RFF and Channel Effects on the Signal}
Without considering the additive noise, the transmitted signal affected by the transmitter's RFF can be expressed as
\begin{equation}
\label{eq6}
x' = {f_T} \otimes x,
\end{equation}
where $ \otimes $ represents the convolution operation and ${f_T}$ is the transmitter's RFF effect. In flat fading channel scenarios, the channel fading is constant across different subcarriers. After being affected by the channel effects, the received signal can be expressed as
\begin{equation}
\label{eq7}
{y_1}^\prime  = \alpha  \cdot x' = \alpha  \cdot {f_T} \otimes x,
\end{equation}
where $\alpha$ represents the flat fading channel effects. The receiver's RFF can also influence the received signal. The signal affected by the receiver's RFF can be expressed as
\begin{equation}
\label{eq8}
y_1 = {f_R} \otimes {y_1}' = \alpha  \cdot {f_R} \otimes {f_T} \otimes x,
\end{equation}
where ${f_R}$ is the receiver's RFF effects. After being transformed into the frequency domain, the final received signal is
\begin{equation}
\label{eq9}
Y_1 = \alpha  \cdot {F_R} \cdot {F_T} \cdot X,
\end{equation}
where ${F_R}$, ${F_T}$, and $X$ are the Fourier transforms (FTs) of ${f_R}$, ${f_t}$, and $x$, respectively.

\subsubsection{RFF Extraction}
To obtain the receiver-agnostic RFF features, a reference device based method is proposed for RFF feature extraction. It is assumed that there is one reference device $T_M$, $I$ transmitters, and $J$ receivers. Since the preamble remains unchanged, the L-STF and L-LTF parts can be utilized to extract RFF features. Here, we use the L-LTF part to demonstrate the RFF extraction process of the RD method. The signal transmitted
\begin{figure}[htbp]
    \centering
    \subfloat{
        \includegraphics[width=4.18cm]{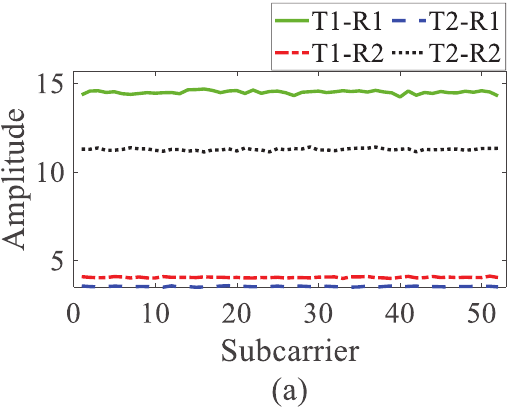}
        }\hfill
    \subfloat{
        \includegraphics[width=4.18cm]{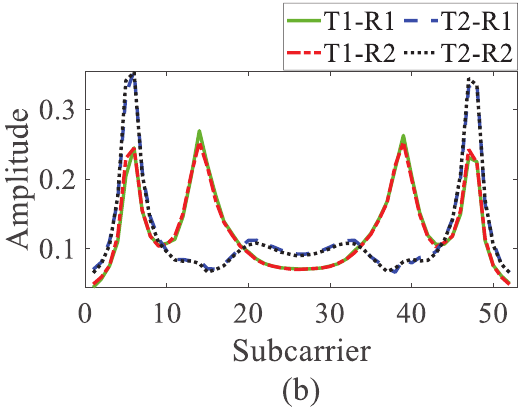}
        }
    \vspace{-0.1cm}
    \caption{The CSI without RFF effects and the extracted RFF features from different WiFi devices, captured by different receivers in the L-LTF part under flat fading channel scenarios in simulations: (a) The CSI without RFF effects and (b) the extracted RFF features.}
    \label{fig4}
    \vspace{-0.3cm}
\end{figure}
by the reference device and received by the $j$-th receiver in the frequency domain can be expressed as
\begin{equation}
\label{eq10}
{Y_{1,{T_M},{R_j}}} = {\alpha _{M,j}} \cdot {F_{{R_j},L}} \cdot {F_{{T_M},L}} \cdot {X_{LLTF}},
\end{equation}
where ${\alpha _{M,j}}$ is the channel effect between the reference device $T_M$ and the $j$-th receiver $R_j$. ${F_{{T_M,L}}}$ and ${F_{{R_j,L}}}$ are the frequency domain RFF effects in the L-LTF part of $T_M$ and $R_j$, respectively. ${X_{LLTF}}$ is the FT of a symbol in the L-LTF part.

The signal transmitted by the $i$-th transmitter and received by the $j$-th receiver in the frequency domain can be expressed as
\begin{equation}
\label{eq11}
{Y_{1,{T_i},{R_j}}} = {\alpha _{i,j}} \cdot {F_{{R_j},L}} \cdot {F_{{T_i},L}} \cdot {X_{LLTF}},
\end{equation}
where ${\alpha_{i,j}}$ represents the channel effect between the $i$-th transmitter $T_i$ and the $j$-th receiver $R_j$, and $F_{T_i,L}$ denotes the frequency domain RFF effects in the L-LTF part of $T_i$.

By dividing the signal in (\ref{eq11}) by the signal in (\ref{eq10}), the receiver-agnostic RFF of the $i$-th device is obtained as
\begin{equation}
\label{eq12}
RF{F_{{i_1}}} = \frac{{{Y_{1,{T_i},{R_j}}}}}{{{Y_{1,{T_M},{R_j}}}}} = \frac{{{\alpha _{i,j}}{F_{{T_i},L}}}}{{{\alpha _{M,j}}{F_{{T_M},L}}}} = {\beta _{i,j,M}}\frac{{{F_{{T_i},L}}}}{{{F_{{T_M},L}}}}.
\end{equation}
The influence of the parameter ${\beta _{i,j,M}}$ can be removed through the normalization process, as it is a constant across different subcarriers.

To obtain additional dimensions of RFF features, the receiver-agnostic RFF features can also be extracted from the L-STF part using the same method. The receiver-agnostic RFF of the $i$-th device extracted from the L-STF part can be expressed as
\begin{equation}
\label{eq13}
RF{F_{{i_2}}} = {\beta _{i,j,M}}\frac{{{F_{{T_i},S}}}}{{{F_{{T_M},S}}}},
\end{equation}
where ${F_{{T_M,S}}}$ and ${F_{{T_i,S}}}$ represent the frequency domain RFF effects in the L-STF part of the reference device $T_M$ and the $i$-th device $T_i$, respectively. The training and classification strategy using both the L-STF and L-LTF parts will be discussed in Section \ref{subsection:5_A}. Note that the extracted RFF features are from the estimated frequency-domain channel state information (CSI), also known as the channel frequency response (CFR). The original CFR contains contributions from the transmitter, the channel, and the receiver. By applying the proposed method to the original CFR, the channel and receiver effects can be removed, allowing the transmitter's RFF features to be extracted.

\begin{table}[h!t]
\scriptsize
   \centering
   \caption{WiFi Modules}
   \vspace{-0.1cm}
    \begin{center}
     \begin{tabular}{||c c c c c c||}
     \hline
     Devices & Dev1 & Dev2 & Dev3 & Dev4 & Dev5\\
     \hline
     Modules & {\makecell[c]{TL-WDR\\4310}} & {\makecell[c]{TL-WDR\\4310}} & {\makecell[c]{TL-WDR\\4310}} & RT-AC66U & RT-N66U\\
     \hline
     Devices & Dev6 & Dev7 & Dev8 & Dev9 & Dev10\\
     \hline
     Modules & RT-N66U & LG-D300 & {\makecell[c]{MAN1K-\\3113NA}} & {\makecell[c]{NETGEAR-\\R6700}} & {\makecell[c]{MOAP\\1200D}} \\
     \hline
     Devices & Dev11 & Dev12 & Dev13 & Dev14 & \\
     \hline
     Modules & {\makecell[c]{LG-\\N595}} & {\makecell[c]{UTT-\\750W}} & {\makecell[c]{GDGEN\\20}} & {\makecell[c]{TL-C2-\\AC750}} & \\
     \hline
    \end{tabular}
    \end{center}
\label{tab2}
\vspace{-0.1cm}
\end{table}
\begin{table}[h!t]
\scriptsize
   \centering
   \caption{USRP Modules}
   \vspace{-0.1cm}
    \begin{center}
     \begin{tabular}{||c c c c c c||}
     \hline
     Receivers & Rx1 & Rx2 & Rx3 & Rx4 & Rx5\\
     \hline
     Modules & {\makecell[c]{USRP \\ N210}}  & {\makecell[c]{LW\\B210}} & {\makecell[c]{LW B205\\ mini-i}} & {\makecell[c]{LW B205\\ mini-i}} & {\makecell[c]{USRP B205\\ mini-i}}\\
     \hline
    \end{tabular}
    \end{center}
\label{tab3}
\vspace{-0.1cm}
\end{table}
\begin{figure}[htbp]
    \centering
    \subfloat{
        \includegraphics[width=4.18cm]{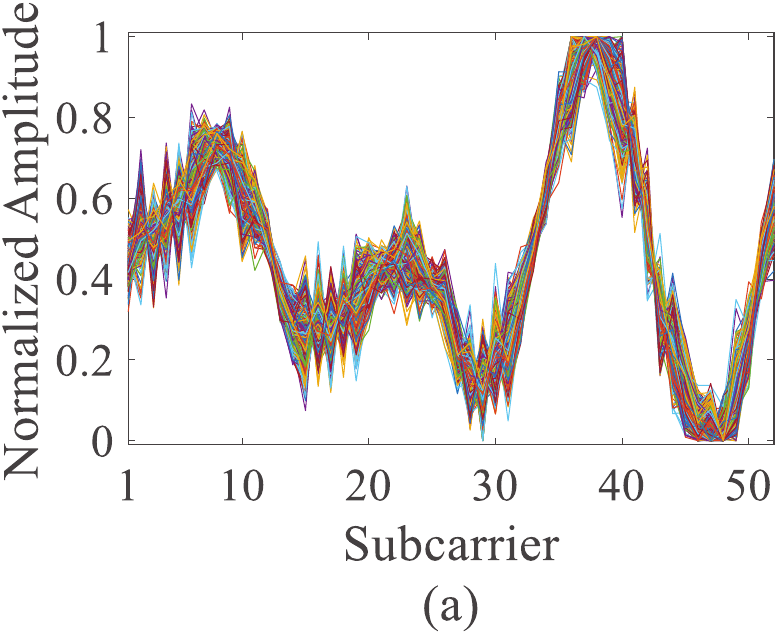}
        }\hfill
    \subfloat{
        \includegraphics[width=4.18cm]{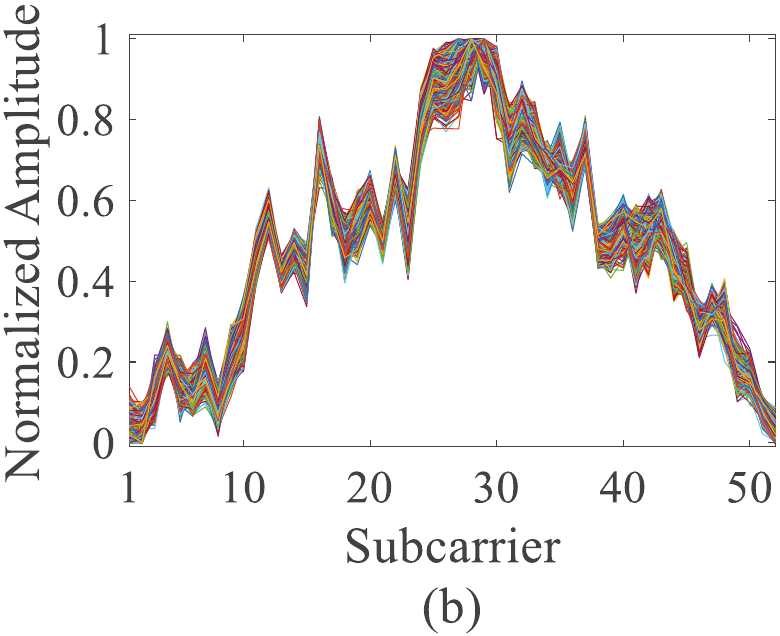}
        }\hfill
    \subfloat{
        \includegraphics[width=4.18cm]{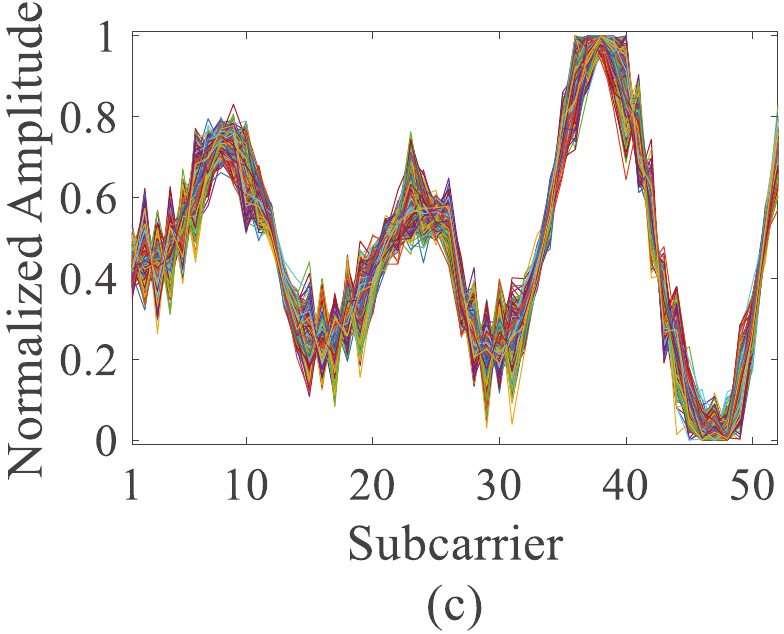}
        }\hfill
    \subfloat{
        \includegraphics[width=4.18cm]{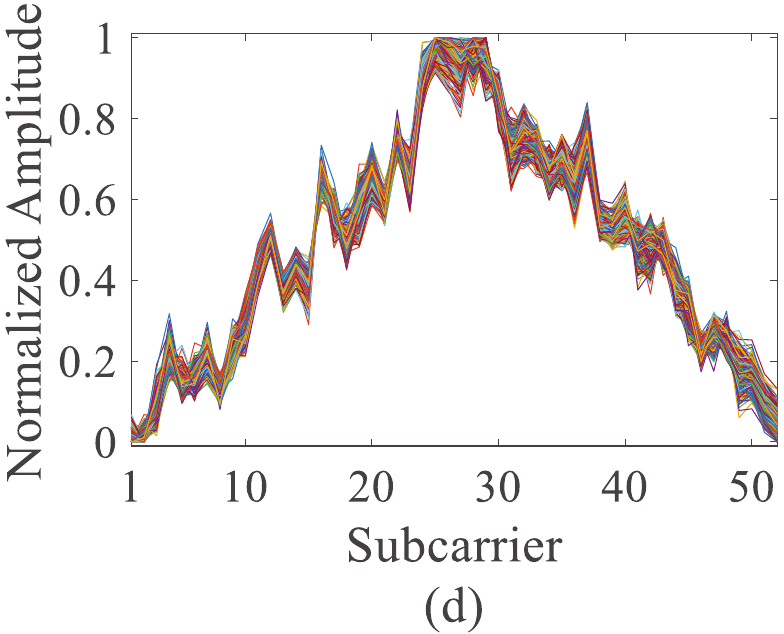}
        }
    \vspace{-0.1cm}
    \caption{Normalized RFF features of different devices, obtained by different receivers in the L-LTF part under the flat fading channel scenario during experiments: (a) Dev5, Rx3; (b) Dev11, Rx3; (c) Dev5, Rx5; and (d) Dev11, Rx5.}
    \label{fig5}
    \vspace{-0.1cm}
\end{figure}

\subsubsection{Evaluation}
Fig. \ref{fig4} illustrates the extracted RFF of devices T1 and T2, captured by different receivers after energy normalization, with a signal-to-noise ratio (SNR) of 40 dB under flat fading channel scenarios in simulations. The distances between different transmitters and receivers are both set to 10 meters in the simulation. Different random seeds are used for the realizations of each transmitter-receiver link. Thus, the channels of different links can be regarded as spatially independent. The CSI of different transmitter-receiver links is shown in Fig. \ref{fig4}(a), where the CSI amplitudes vary across links. As illustrated in Fig. \ref{fig4}, the extracted RFF features of each transmitter remain consistent despite spatial channel variations and receiver differences. Therefore, the proposed RD method demonstrates robustness to spatial channel variations and receiver differences.

Table \ref{tab2} presents the types of WiFi modules, while Table \ref{tab3} shows the types of USRP modules employed for signal receiving. Fig. \ref{fig5} shows the normalized RFF features of Dev5 and Dev11, obtained by Rx3 and Rx5 from the L-LTF part in a wired scenario \cite{gaskin2023deep, gaskin2022tweak}.
\begin{figure}[htbp]
    \centering
    \subfloat{
        \includegraphics[width=4.18cm]{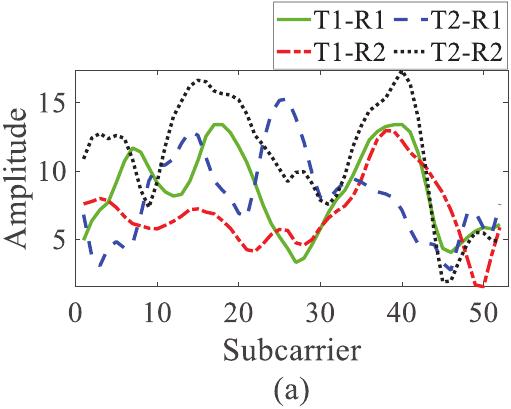}
        }\hfill
    \subfloat{
        \includegraphics[width=4.18cm]{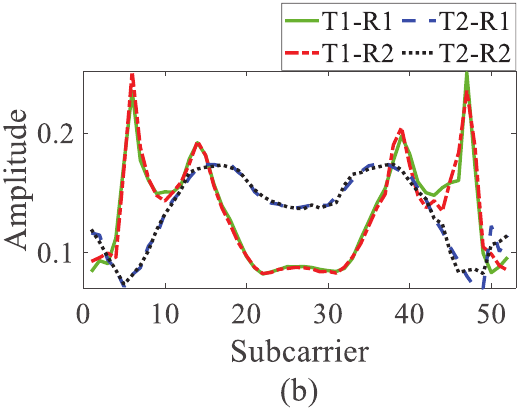}
        }
    \vspace{-0.1cm}
    \caption{The CSI without RFF effects and the extracted RFF features from different devices, captured by different receivers using the HL method under frequency-selective fading channel scenarios in simulations: (a) The CSI without RFF effects and (b) the extracted RFF features.}
    \label{fig6}
    \vspace{-0.3cm}
\end{figure}
It can be observed that the extracted RFF features of different devices vary. However, for each individual device, the RFF features extracted by different receivers remain similar. Therefore, the proposed RD method can extract receiver-agnostic RFF features under the flat fading channel scenario.

\subsection{RFF Extraction under Frequency-selective Fading Channel Scenarios}
\subsubsection{RFF and Channel Effects on the Signal}
Under frequency-selective fading channel scenarios, the transmitted signal has the same form as in (\ref{eq6}). The received signal, affected by frequency-selective channel effects, can be expressed as
\begin{equation}
\label{eq14}
{y_2}' = h \otimes x' = h \otimes {f_T} \otimes x,
\end{equation}
where $h$ denotes the frequency-selective channel effects. The signal ${y_2}'$, affected by the receiver's RFF, can be expressed as
\begin{equation}
\label{eq15}
{y_2} = {f_R} \otimes {y_2}' = h \otimes {f_R} \otimes {f_T} \otimes x.
\end{equation}
The signal ${y_2}$ can be transformed into the frequency domain as
\begin{equation}
\label{eq16}
{Y_2} = H \cdot {F_R} \cdot {F_T} \cdot X,
\end{equation}
where $H$ is the FT of $h$.

\subsubsection{RFF Extraction}
A division-based receiver-agnostic RFF identification method is proposed to eliminate both the channel effects and the receiver's RFF effects. In the L-LTF part, the signal transmitted by the $i$-th transmitter $T_i$ and received by the $j$-th receiver $R_j$ in the frequency domain can be expressed as
\begin{equation}
\label{eq17}
{Y_{{2_L},{T_i},{R_j}}} = H \cdot {F_{{R_j},L}} \cdot {F_{{T_i},L}} \cdot {X_{L}}.
\end{equation}
In the HT-LTF part, the signal transmitted by $T_i$ and received by $R_j$ in the frequency domain can be expressed as
\begin{equation}
\label{eq18}
{Y_{{2_H},{T_i},{R_j}}} = H \cdot {F_{{R_j},H}} \cdot {F_{{T_i},H}} \cdot {X_{H}}.
\end{equation}
In this paper, we assumed that the RFFs across HT-LTF and L-LTF parts are different of the device to be classified, while the RFFs across HT-LTF and L-LTF parts are similar of the receivers. Therefore, we have
\begin{equation}
\label{eq19}
{F_{{R_j},H}} = {F_{{R_j},L}},
\end{equation}
\begin{equation}
\label{eq20}
{F_{{T_i},H}} \ne {F_{{T_i},L}}.
\end{equation}

\begin{figure}[htbp]
    \centering
    \subfloat{
        \includegraphics[width=4.18cm]{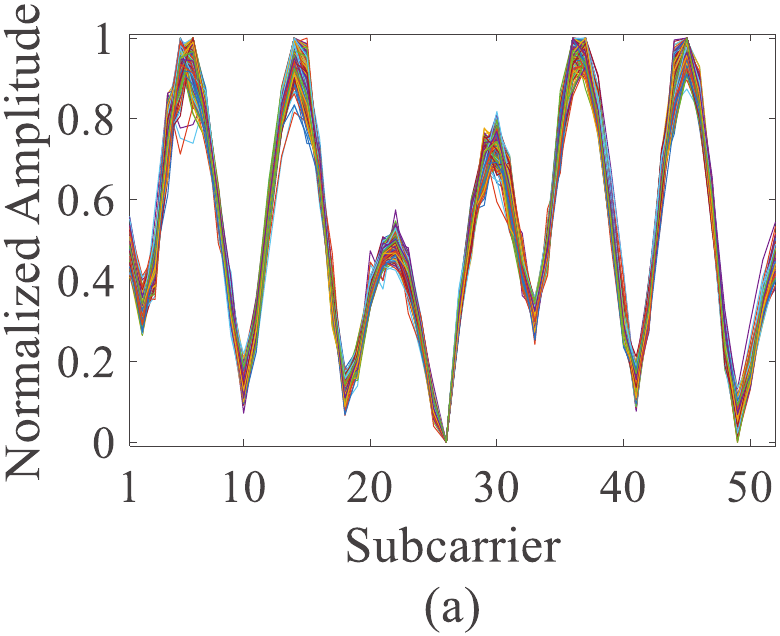}
        }\hfill
    \subfloat{
        \includegraphics[width=4.18cm]{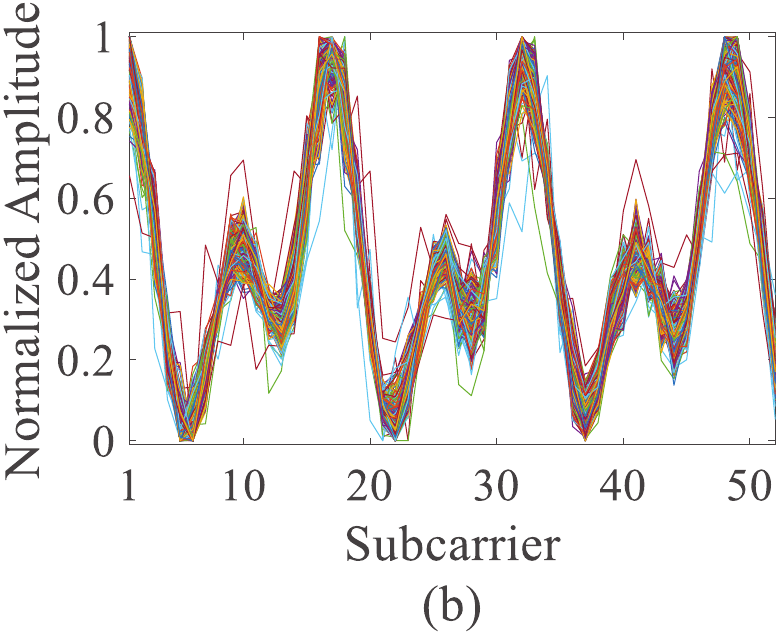}
        }\hfill
    \subfloat{
        \includegraphics[width=4.18cm]{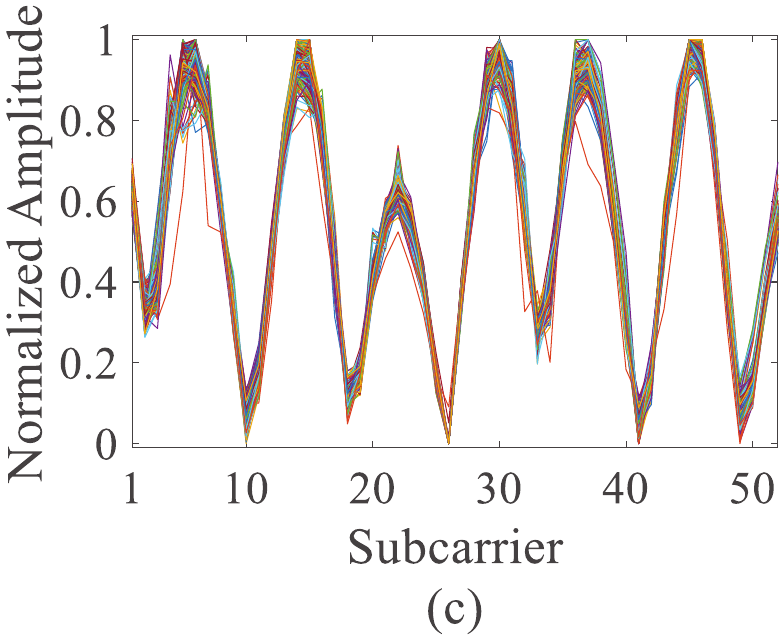}
        }\hfill
    \subfloat{
        \includegraphics[width=4.18cm]{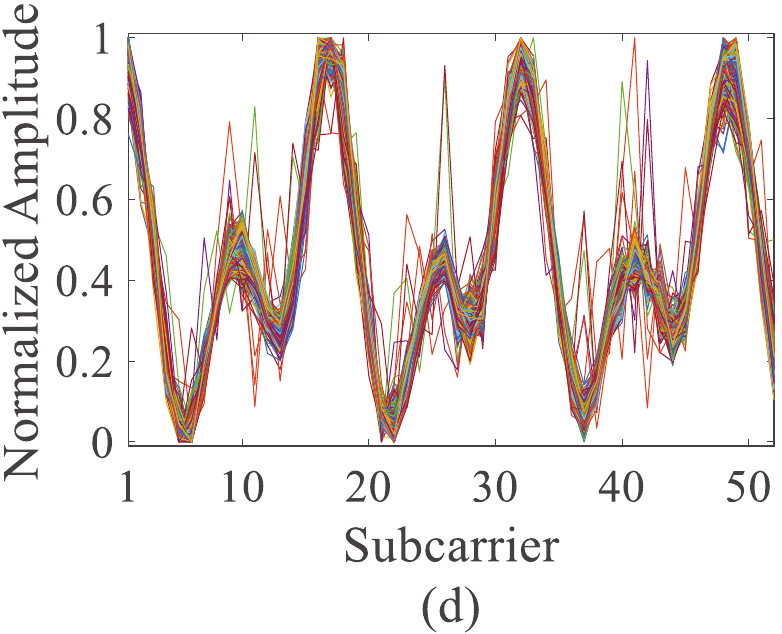}
        }
    \vspace{-0.1cm}
    \caption{Normalized RFF features of different devices, obtained by different receivers using the HL method under the frequency-selective fading channel scenario during experiments: (a) Dev6, Rx3; (b) Dev13, Rx3; (c) Dev6, Rx5; and (d) Dev13, Rx5.}
    \label{fig7}
    \vspace{-0.3cm}
\end{figure}
The receiver-agnostic RFF of $T_i$ can be obtained by dividing the signal in (\ref{eq18}) by the signal in (\ref{eq17}) in the frequency domain as
\begin{equation}
\label{eq21}
\begin{array}{l}
RFF_{i_2}=\frac{Y_{2_H,T_i,R_j}}{Y_{2_L,T_i,R_j}}=\frac{H\cdot F_{R_j,H}\cdot F_{T_i,H}\cdot X_H}{H\cdot F_{R_j,L}\cdot F_{T_i,L}\cdot X_L}=\gamma _{HL}\frac{F_{T_i,H}}{F_{T_i,L}},
\end{array}
\end{equation}
where ${\gamma _{HL}}$ is constant across different transmissions because ${{X_H}}$ and ${{X_L}}$ are invariant.
\subsubsection{Evaluation}
Fig. \ref{fig6} illustrates the extracted RFFs of T1 and T2, obtained by different receivers after energy normalization, under frequency-selective fading channel scenarios with an SNR of 40 dB in simulations. Similar to the simulation setup in Fig. \ref{fig4}, the distances between different transmitters and receivers are both set to 10 meters. Different random seeds are used for the realizations of different transmitter-receiver links. The CSI of different transmitter-receiver links is shown in Fig. \ref{fig6}(a), where the CSI amplitudes also differ across links. As shown in Fig. \ref{fig6}(b), despite  spatial channel variations and receiver differences, the extracted RFF features of each transmitter remain consistent. Thus, the proposed HL method can mitigate the adverse effects of spatial channel variations and receiver differences.

Fig. \ref{fig7} shows the normalized RFF features of Dev6 and Dev13, obtained by receivers Rx3 and Rx5 during  experiments. As shown in Fig. \ref{fig7}, the extracted RFF features of different devices are distinguishable. Moreover, the RFF features of each device, extracted by different receivers, remain similar. Thus, the proposed HL method can extract receiver-agnostic RFF features under frequency-selective fading channel scenarios.

\section{Experimental Evaluation}
\label{section:5}

\subsection{Experimental Setting}
\label{subsection:5_A}

\begin{table}[h!t]
\scriptsize
   \centering
   \caption{WiFi Modules in Different Evaluation Sets}
   \vspace{-0.1cm}
    \begin{center}
     \begin{tabular}{||c c c c||}
     \hline
     Devices & WiFi Module & Evaluation Set S1 & Evaluation Set S2 \\
     \hline
     Dev1 &TL-WDR4310&  & \Checkmark\\
     \hline
     Dev2 &TL-WDR4310& \CheckmarkBold(Reference Device) & \Checkmark\\
     \hline
     Dev3 &TL-WDR4310&  & \Checkmark\\
     \hline
     Dev4 &RT-AC66U& \Checkmark & \\
     \hline
     Dev5 &RT-N66U& \Checkmark & \\
     \hline
     Dev6 &RT-N66U& \Checkmark & \Checkmark\\
     \hline
     Dev7 &LG-D300& \Checkmark & \Checkmark\\
     \hline
     Dev8 &MAN1K-3113NA& \Checkmark & \Checkmark\\
     \hline
     Dev9 &NETGEAR-R6700& & \Checkmark\\
     \hline
     Dev10 &MOAP1200D & \Checkmark &\\
     \hline
     Dev11 &LG-N595& \Checkmark & \\
     \hline
     Dev12 &UTT-750W& \Checkmark & \Checkmark\\
     \hline
     Dev13 &GDGEN20& \Checkmark & \Checkmark\\
     \hline
     Dev14 &TL-C2-AC750& \Checkmark & \Checkmark\\
     \hline
    \end{tabular}
    \end{center}
\label{tab4}
\vspace{-0.3cm}
\end{table}
\begin{table}[h!t]
\scriptsize
   \centering
   \caption{Training Set}
   \vspace{-0.1cm}
    \begin{center}
     \begin{tabular}{||c c||}
     \hline
     Training Set & USRP \\
     \hline
     Set1 &  Rx2, Rx3\\
     \hline
     Set2 &  Rx2, Rx3, Rx4\\
     \hline
     Set3 &  Rx2, Rx3, Rx4, Rx5\\
     \hline
     Set4 &  Rx1, Rx2\\
     \hline
     Set5 &  Rx1, Rx2, Rx4\\
     \hline
     Set6 &  Rx1, Rx2, Rx4, Rx5\\
     \hline
    \end{tabular}
    \end{center}
\label{tab5}
\vspace{-0.3cm}
\end{table}
\subsubsection{Signal Collection}
First, the cable wire is used to connect the WiFi module directly to the universal software radio peripheral (USRP) for evaluating the performance of the proposed method in the flat fading channel scenario \cite{gaskin2023deep, gaskin2022tweak}. The sampling rate of the USRP is set to 20 Msps. A radio frequency attenuator with 30 dB attenuation is employed to protect the USRP. The carrier frequency of the WiFi devices is set to 5.745 GHz, corresponding to the 149th WiFi channel. Second, to evaluate the performance of the proposed method under the frequency-selective fading channel scenarios, experiments are conducted in a typical packed indoor office and a corridor with dynamic passersby. Signals are collected in the packed indoor office under static line-of-sight (LOS), static non-line-of-sight (NLOS), and mobile scenarios, and in the corridor under the static LOS scenario. Table \ref{tab4} presents the types of WiFi modules used in the different experiments, and Table \ref{tab3} shows the types of USRP modules employed for signal receiving.

As illustrate in Table \ref{tab4}, evaluation set S1 consist of one reference device and ten WiFi devices for classification. Evaluation set S2 also employs ten WiFi devices for classification, but without a reference device. To train neural networks with signals received from multiple transmitters, six distinct training sets are defined, as shown in Table \ref{tab5}.
\subsubsection{Neural Network for Classification}
Since neural networks are widely used in RFF identification, we utilize the InceptionTime neural network \cite{ismail2020inceptiontime}, which has been shown to achieve high classification accuracy \cite{yang2023led}. Similar to the approach in \cite{yang2023led}, soft labeling is employed to prevent overfitting, and L2 regularization is incorporated to improve the model's generalization performance.

\begin{figure}[htbp]%
    \centering
    \subfloat{
        \includegraphics[width=4.18cm]{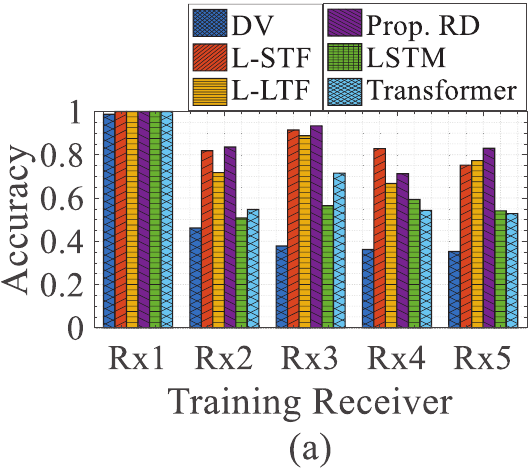}
        }\hfill
    \subfloat{
        \includegraphics[width=4.18cm]{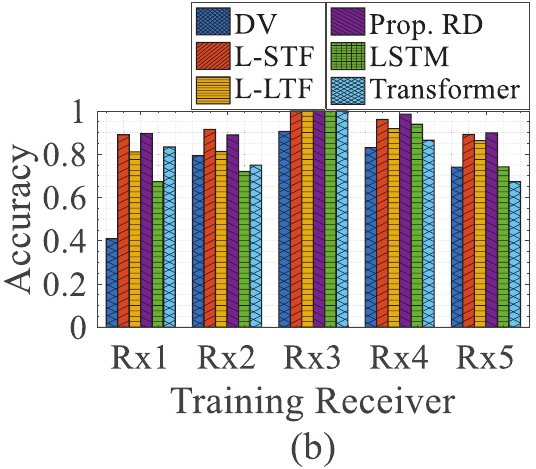}
        }
    \vspace{-0.1cm}
    \caption{Classification accuracy with training on a single receiver under the flat fading channel scenario: (a) Tested on Rx1 and (b) tested on Rx3.}
    \label{fig8}
    \vspace{-0.3cm}
\end{figure}

\subsubsection{Training of Neural Network}
To train the neural network, the label smoothing factor, number of epochs, batch size, learning rate, and L2 regularization are set to 0.1, 50, 64, 0.001, and 0.1, respectively. The model with the lowest loss value on the validation set is saved as the best model. The classification accuracy is averaged using the Monte Carlo multiple validation method. Under the flat fading channel scenario, the RFF features extracted from the L-STF and L-LTF parts are used to train two separate neural networks during the training process. In the testing process, the softmax outputs of the two individual networks are combined, and the category with the highest value is considered the predicted result \cite{qiu2022signal}.
\subsubsection{The Evaluation Metric}
We use classification accuracy as the evaluation metric to compare the classification performance of different methods. Classification accuracy is defined as the ratio of correctly classified samples to the total number of testing samples. To assess the classification performance, the existing division-based RFF extraction method, denoted as the DV method \cite{chen2022radioDSRC, xing2022design}, and the GAN-based two-stage supervised learning framework, denoted as the GAN-RXA method \cite{zhao2023gan}, are utilized as benchmarks in experiments.

\subsection{Classification under the Flat Fading Channel Scenario}
In this subsection, we evaluate the classification performance of different methods across varying receivers in the flat fading channel scenario using evaluation set S1.

\subsubsection{Single Receiver Condition}
Fig. \ref{fig8} shows the classification accuracy using only one receiver for training. The test sets are collected by Rx1 and Rx3, as shown in Fig. \ref{fig8}(a) and Fig. \ref{fig8}(b), respectively. The GAN-RXA method, which will be evaluated in the multiple receivers condition, is not included here, as it requires the training set to be sourced from multiple receivers. In Fig. \ref{fig8}, the L-STF and L-LTF methods represent the RFF features extracted solely from the L-STF or L-LTF part, respectively, used for device classification.

It can be observed from Fig. \ref{fig8} that the accuracy of all evaluated methods decreases when the training and testing receivers differ.
\begin{figure}[htbp]
    \centering
    \subfloat{
        \includegraphics[width=4.18cm]{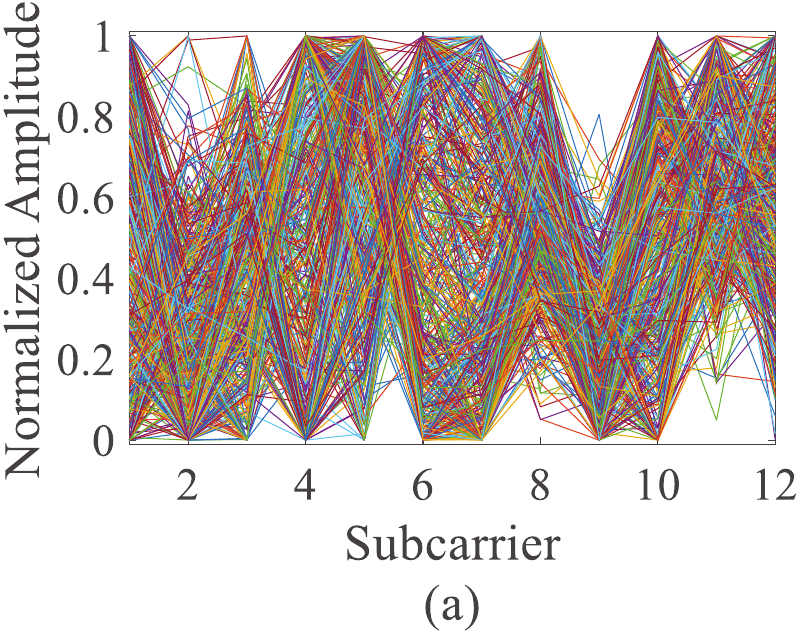}
        }\hfill
    \subfloat{
        \includegraphics[width=4.18cm]{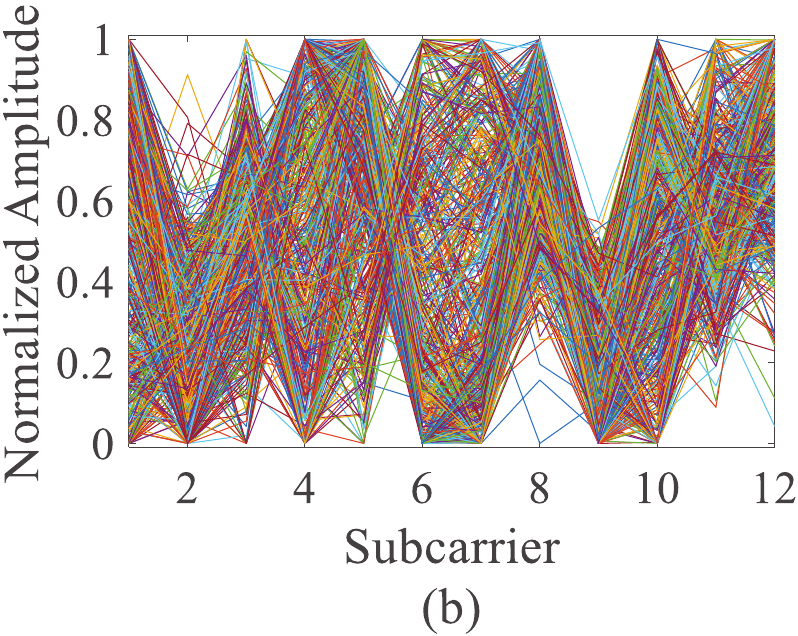}
        }
    \vspace{-0.1cm}
    \caption{Normalized RFF features of Dev5 obtained by different receivers using the DV method \cite{chen2022radioDSRC, xing2022design}: (a) Rx3 and (b) Rx5.}
    \label{fig9}
    \vspace{-0.3cm}
\end{figure}
\begin{figure}
\centering
\includegraphics[width=7cm]{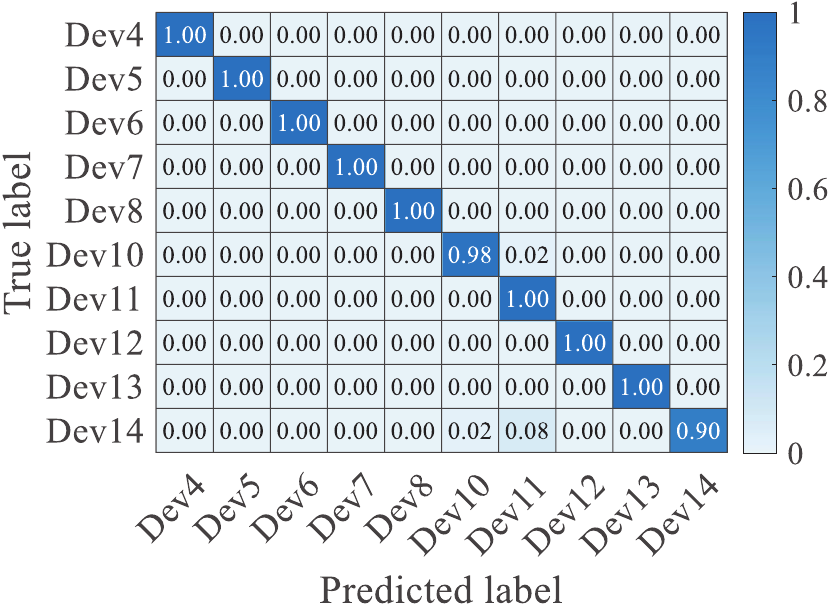}
\vspace{-0.1cm}
\caption{The confusion matrix for the proposed RD method, with training and test sets collected by Rx4 and Rx3, respectively.}
\label{fig10}
\vspace{-0.3cm}
\end{figure}
Meanwhile, the proposed RD method achieves higher classification accuracy compared to the L-STF method, L-LTF method, and the DV method in most cases. The classification accuracies exceed 80\% in most cases, indicating that the proposed method is practical for use. In contrast, the change in receivers significantly impacts the classification accuracy of the DV method, especially when training on Rx5 and testing on Rx1, where the classification accuracy drops to merely 35.25\%. This is because the RFF features in the L-STF and L-LTF parts are similar, and thus, the RFF features extracted by dividing the L-STF part from the L-LTF part in the frequency domain are more susceptible to noise.

The noise sensitive RFF features extracted by the DV method are illustrated in Fig. \ref{fig9}, which shows the RFF features of Dev5 received by different receivers. By comparing Fig. \ref{fig9}(a) and Fig. \ref{fig9}(b) with Fig. \ref{fig5}(a) and Fig. \ref{fig5}(c), it can be observed that the RFF features extracted by the DV method are less stable than those extracted by the RD method. This is because the proposed RD method extracts RFF features by dividing the preamble of the device to be classified from that of the reference device. Since the RFFs of the reference device and the device to be classified are different, the RFF features extracted by the RD method are more stable, which is beneficial for classification. Therefore, the stacking process required in \cite{chen2022radioDSRC, xing2022design} for denoising is not necessary in the proposed RD method.

The neural network structures of LSTM and Transformer used in \cite{shen2023towards} are adopted for the RFF identification task, and the corresponding classification accuracies are shown in Fig. \ref{fig8}. As illustrated in Fig. \ref{fig8}, the classification accuracy of InceptionTime is consistently higher than that of LSTM and Transformer when the receiver changes. One plausible reason for this is that because CSI-based feature classification resembles a time-series classification task, for which InceptionTime serves as a more suitable and powerful time-series classifier.
\begin{figure}[htbp]
    \centering
    \subfloat{
        \includegraphics[width=4.18cm]{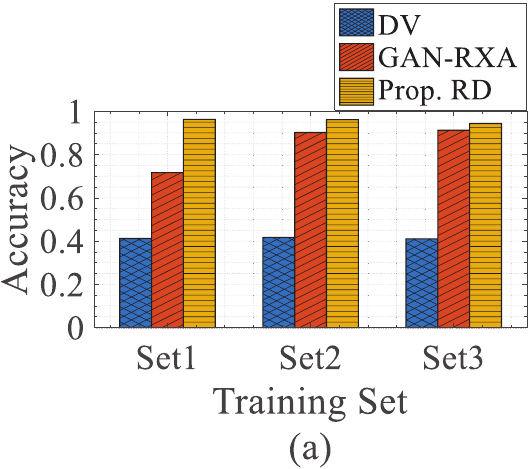}
        }\hfill
    \subfloat{
        \includegraphics[width=4.18cm]{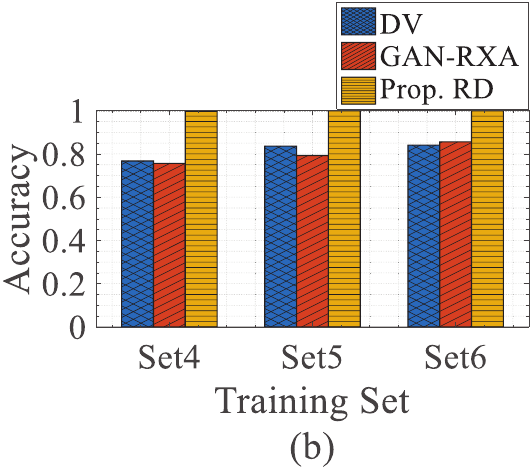}
        }
    \vspace{-0.1cm}
    \caption{Classification accuracy with training on multiple transmitters under the flat fading channel scenario: (a) Tested on Rx1 and (b) tested on Rx3.}
    \label{fig11}
    \vspace{-0.3cm}
\end{figure}
Fig. \ref{fig10} shows the confusion matrix for the proposed RD method. The training and test sets are collected by Rx4 and Rx3, respectively. It can be observed from Fig. \ref{fig10} that most devices achieve 100\% classification accuracy. This is because the RFF features extracted by the RD method are resilient to the impacts of receivers.

\subsubsection{Multiple Receivers Condition}

The classification performance is evaluated when signals collected by multiple receivers are used for training, and the results are shown in Fig. \ref{fig11}. It can be observed from Fig. \ref{fig11} that the proposed RD method consistently achieves the highest classification accuracy among all the methods. The classification accuracy of the RD method exceeds that of the GAN-RXA method by 3.11\% to 24.66\%.

\subsubsection{The Influence of the Reference Device Selection}
\begin{figure}
\centering
\includegraphics[width=5cm]{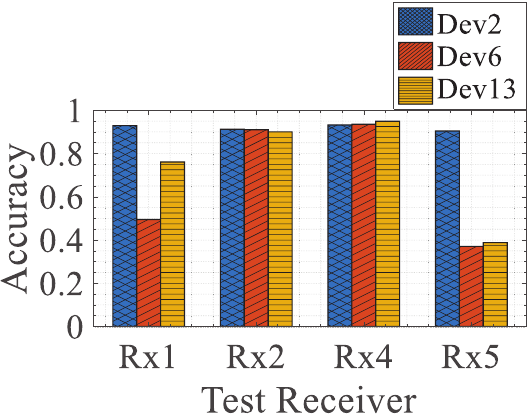}
\vspace{-0.1cm}
\caption{Classification accuracies with training on Rx3 using different reference devices and testing on various receivers under the flat fading channel scenario.}
\label{fig12}
\vspace{-0.3cm}
\end{figure}

To evaluate the impact of reference device selection, different reference devices are employed, and the corresponding classification performance is illustrated in Fig. \ref{fig12}. The evaluation set is still S1. Dev2, Dev6, and Dev13 are selected as reference devices, while the remaining eight devices are used for classification. As shown in Fig. \ref{fig12}, the classification accuracy remains above 90\% when training on Rx3, testing on different receivers, and using Dev2 as the reference device. In contrast, when Dev6 or Dev13 are selected as the reference device, the classification accuracy is low in some cases, such as training on Rx5 and testing on Rx1.

\begin{figure}[htbp]
    \centering
    \subfloat{
        \includegraphics[width=4cm]{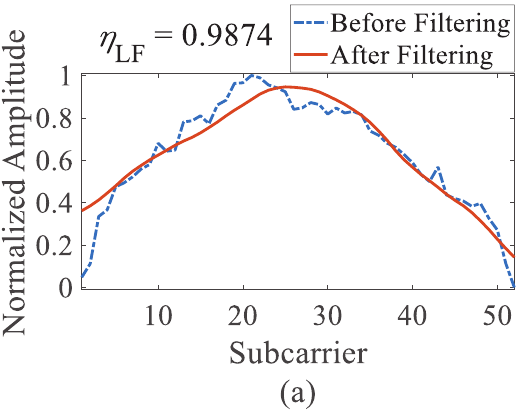}
        }\hfill
    \subfloat{
        \includegraphics[width=4cm]{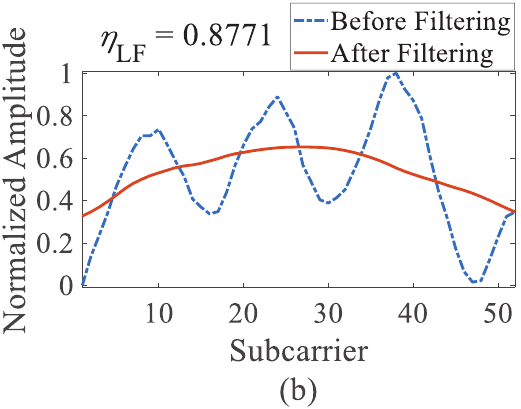}
        }\hfill
    \subfloat{
        \includegraphics[width=4cm]{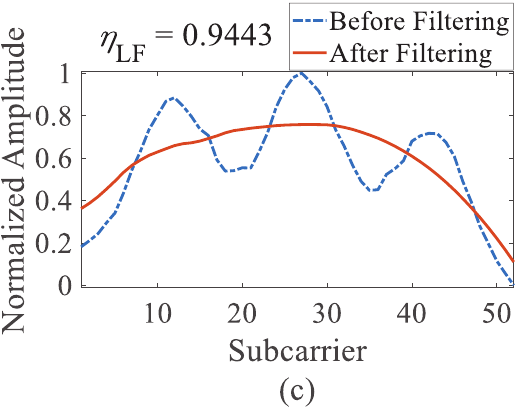}
        }
    \vspace{-0.1cm}
    \caption{The CSI from the L-LTF part of different reference devices before and after wavelet filtering: (a) Dev2, (b) Dev6, and (c) Dev13.}
    \label{fig13}
    \vspace{-0.3cm}
\end{figure}
\begin{figure}
\centering
\includegraphics[width=5cm]{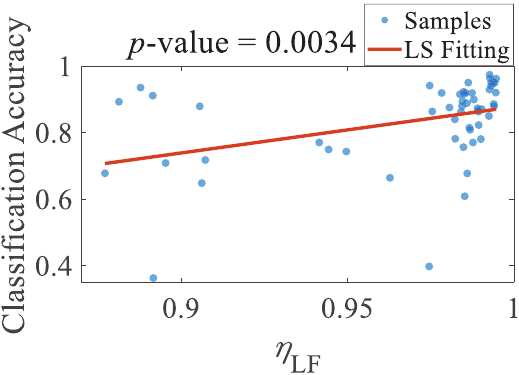}
\vspace{-0.1cm}
\caption{The correlation between the metric $\eta _{\mathrm{LF}}$ and classification accuracy when the receiver changes.}
\label{fig14}
\vspace{-0.3cm}
\end{figure}
Since the classification accuracy depends on the reference device, a metric should be designed for reference device selection. After normalization, the amplitude of CSI is decomposed using the Daubechies order-4 (db4) wavelet into four layers and reconstructed from the approximation coefficients in the fourth layer to obtain the low-frequency component of the normalized CSI. The energies of the normalized CSI before and after wavelet filtering are denoted by $E_B$ and $E_A$, respectively. Then, a metric for selecting the reference device can be expressed as
\begin{equation}
\label{eq22}
\eta _{\mathrm{LF}}=\frac{E_A}{E_B}.
\end{equation}
The normalized CSI of different reference devices collected by Rx3, both before and after wavelet filtering, and the corresponding $\eta _{\mathrm{LF}}$ are shown in Fig. \ref{fig13}. As illustrated in Fig. \ref{fig13}, the reference device Dev2, which achieves the highest classification accuracy on average, also exhibits the highest $\eta _{\mathrm{LF}}$.

Fig. \ref{fig14} shows the $\eta _{\mathrm{LF}}$ values of different reference devices and the corresponding classification accuracies when tested on different receivers. Based on these two-dimensional samples, the $p$-value between $\eta _{\mathrm{LF}}$ and classification accuracy is calculated as 0.0034. Since the $p$-value is much smaller than 0.05, there is a strong correlation between $\eta _{\mathrm{LF}}$ and classification accuracy. By applying least-squares (LS) fitting to these samples, it can be observed that and classification accuracy are positively correlated. Therefore, $\eta _{\mathrm{LF}}$ can be utilized as a metric for reference device selection.

\begin{figure}[htbp]
    \centering
    \subfloat{
        \includegraphics[width=4.18cm]{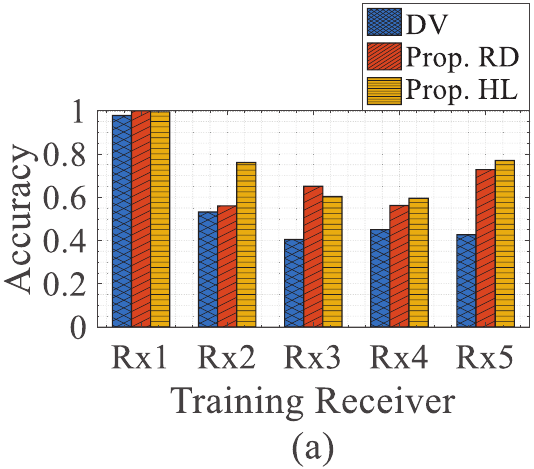}
        }\hfill
    \subfloat{
        \includegraphics[width=4.18cm]{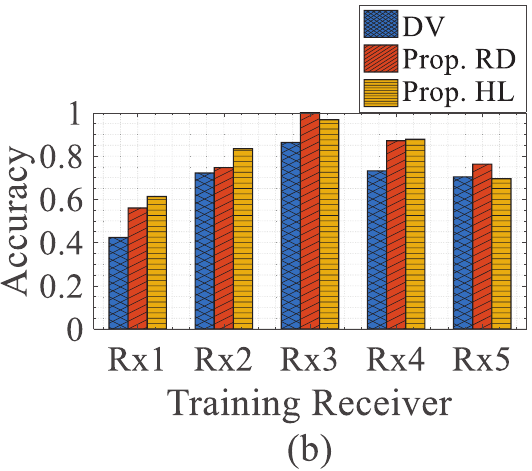}
        }
    \vspace{-0.1cm}
    \caption{Classification accuracy with training on the evaluation set S2 under the flat fading channel scenario: (a) Tested on Rx1 and (b) tested on Rx3.}
    \label{fig15}
    \vspace{-0.3cm}
\end{figure}
\begin{figure}[htbp]
    \centering
    \subfloat{
        \includegraphics[width=4.18cm]{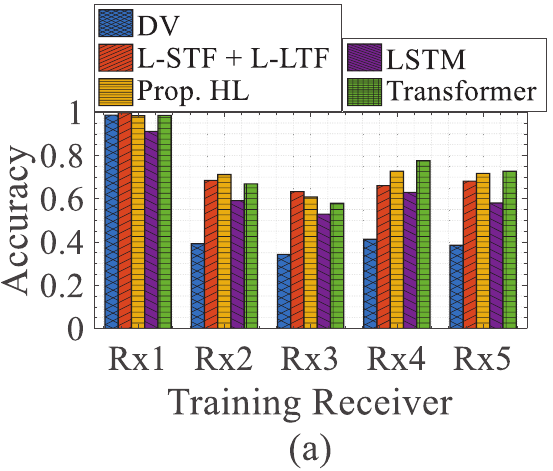}
        }\hfill
    \subfloat{
        \includegraphics[width=4.18cm]{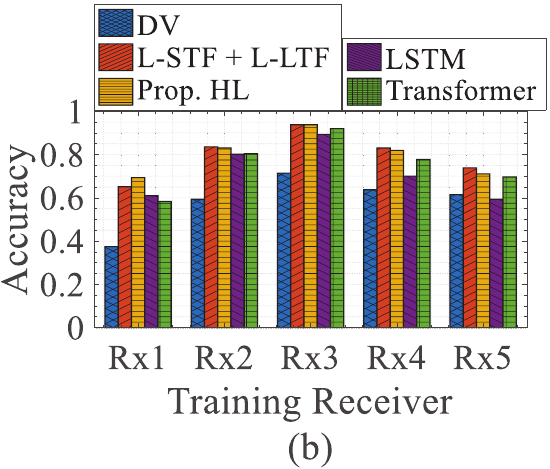}
        }
    \vspace{-0.1cm}
    \caption{Classification accuracy with training on a single receiver under the static LOS scenario: (a) Tested on Rx1 and (b) tested on Rx3.}
    \label{fig16}
    \vspace{-0.3cm}
\end{figure}

\subsubsection{Comparison between the RD and HL Methods}
Since the HL method is based on the HT-MF frame, we evaluate the classification performance using test set S2, where the HT-MF frame appears more frequently. As shown in Fig. \ref{fig15}, the classification accuracies of the proposed RD and HL methods are similar. However, since HT-MF signals, which are rare in some devices, are required in the HL method, the RD method is more universal, as both non-HT and HT-MF signals are suitable for RFF feature extraction in the RD method.

\subsection{Classification under Frequency-selective Fading Channel Scenarios}
In this subsection, we evaluate the classification performance of different methods under receiver differences in frequency-selective fading channel scenarios using evaluation set S2. All devices in evaluation set S2, except Dev2 and Dev3, are used in the experiments conducted under the static corridor scenario.

\subsubsection{Classification under the Same Channel Scenario}
\begin{figure}[htbp]
    \centering
    \subfloat{
        \includegraphics[width=4.18cm]{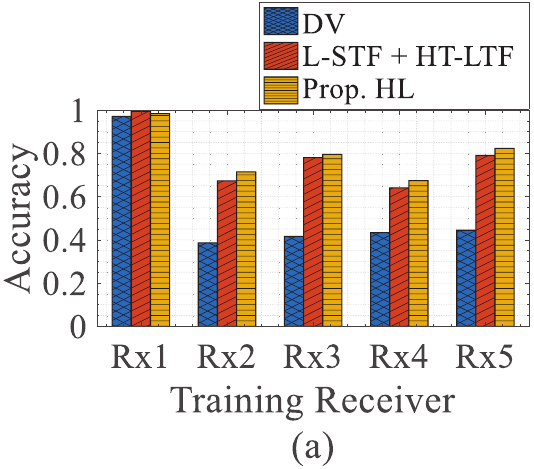}
        }\hfill
    \subfloat{
        \includegraphics[width=4.18cm]{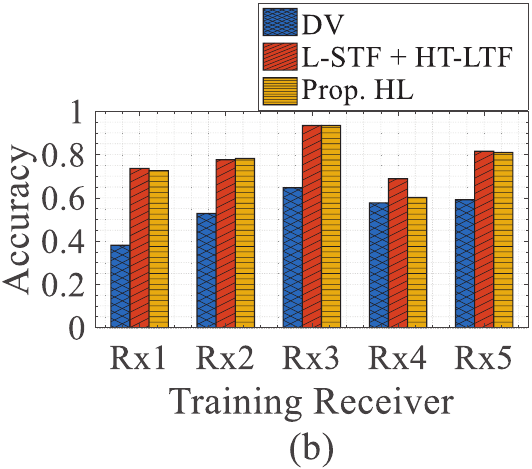}
        }
    \vspace{-0.1cm}
    \caption{Classification accuracy with training on a single receiver in the static NLOS scenario: (a) Tested on Rx1 and (b) tested on Rx3.}
    \label{fig17}
    \vspace{-0.3cm}
\end{figure}

\begin{figure}[htbp]
    \centering
    \subfloat{
        \includegraphics[width=4.18cm]{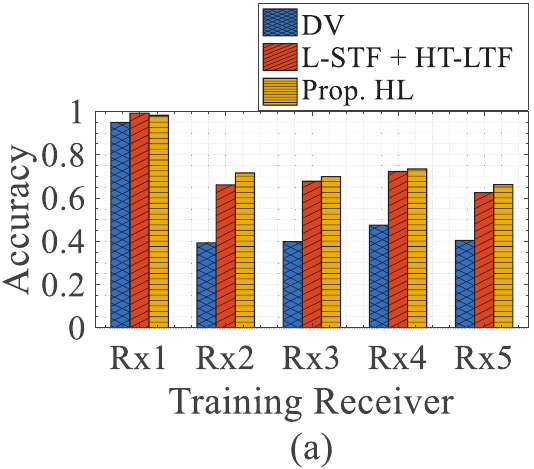}
        }\hfill
    \subfloat{
        \includegraphics[width=4.18cm]{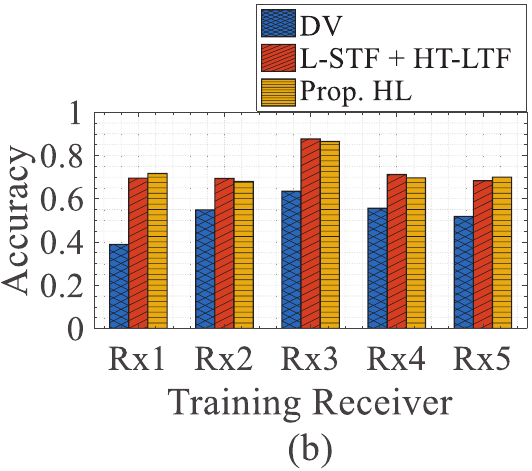}
        }
    \vspace{-0.1cm}
    \caption{Classification accuracy with training on a single receiver in the mobile scenario: (a) Tested on Rx1 and (b) tested on Rx3.}
    \label{fig18}
    \vspace{-0cm}
\end{figure}

Fig. \ref{fig16}, Fig. \ref{fig17}, and Fig. \ref{fig18} show the classification accuracy of the DV method and the proposed HL method when the training set is collected from a single receiver in the packed indoor office under the static LOS, static NLOS, and mobile scenarios, respectively. As shown in these figures, classification accuracy decreases when the receiver changes. However, the classification accuracy of the proposed HL method is consistently higher than that of the DV method by 2.44\% to 37.95\% when the training and test sets are collected from different receivers, demonstrating a significant classification accuracy improvement in most cases. This is because the proposed HL method extracts higher-dimensional RFF features, i.e., 52 subcarriers, compared to the DV method, i.e., 12 subcarriers \cite{chen2022radioDSRC, xing2022design}.
\begin{figure}[htbp]
    \centering
    \subfloat{
        \includegraphics[width=4.18cm]{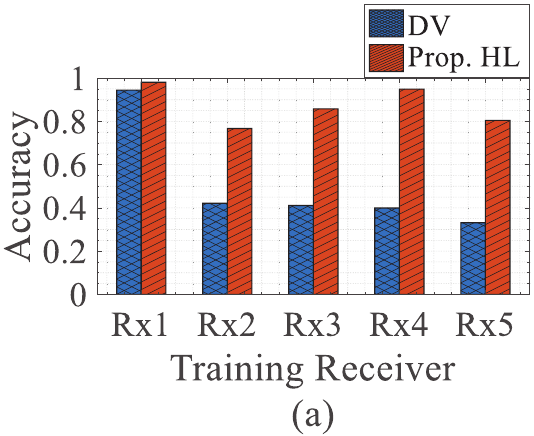}
        }\hfill
    \subfloat{
        \includegraphics[width=4.18cm]{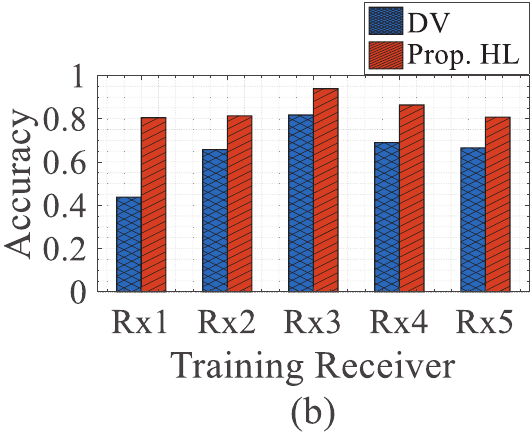}
        }
    \vspace{-0.1cm}
    \caption{Classification accuracy with training on a single receiver under the static corridor scenario: (a) Tested on Rx1 and (b) tested on Rx3.}
    \label{fig19}
    \vspace{-0.3cm}
\end{figure}

The classification accuracies of InceptionTime, LSTM, and Transformer are compared under the static LOS scenario, as shown in Fig. \ref{fig16}. As illustrated in Fig. \ref{fig16}, the classification accuracies of InceptionTime and Transformer are comparable, whereas LSTM achieves the lowest accuracy. Therefore, as a strong time-series classifier, InceptionTime is well suited for RFF identification based on features extracted from CSI amplitude.

\begin{figure}[htbp]
    \centering
    \subfloat{
        \includegraphics[width=4.18cm]{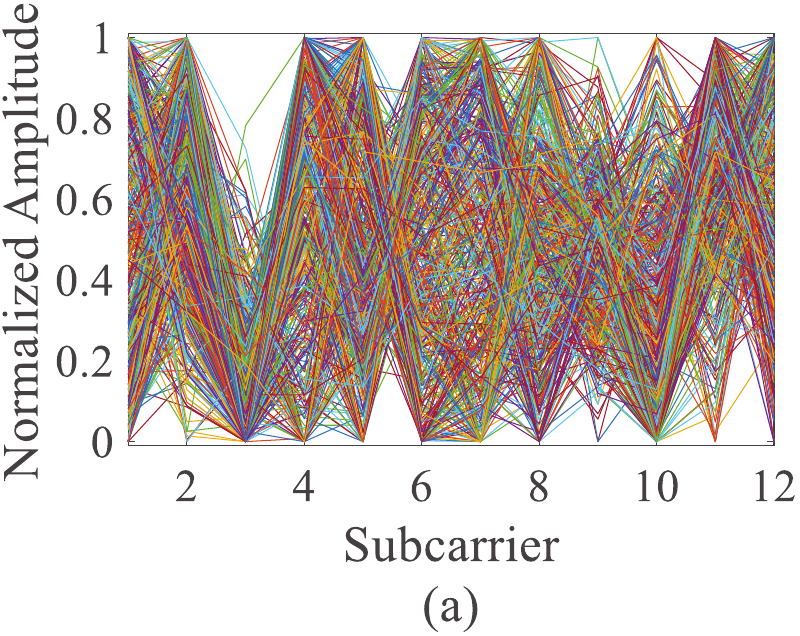}
        }\hfill
    \subfloat{
        \includegraphics[width=4.18cm]{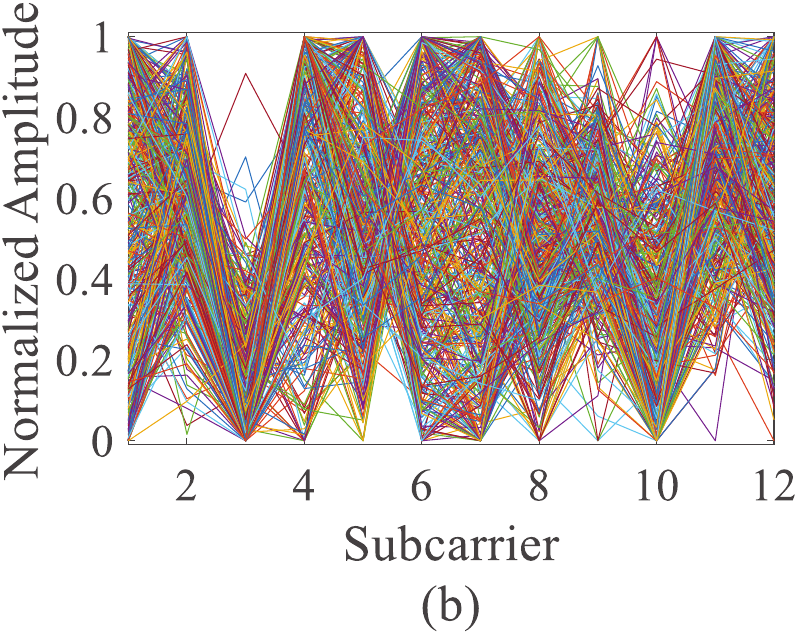}
        }
    \vspace{-0.1cm}
    \caption{Normalized RFF features of Dev13 obtained by different receivers using the DV method \cite{chen2022radioDSRC, xing2022design}: (a) Rx3 and (b) Rx5.}
    \label{fig20}
    \vspace{-0.3cm}
\end{figure}
\begin{figure}
\centering
\includegraphics[width=7cm]{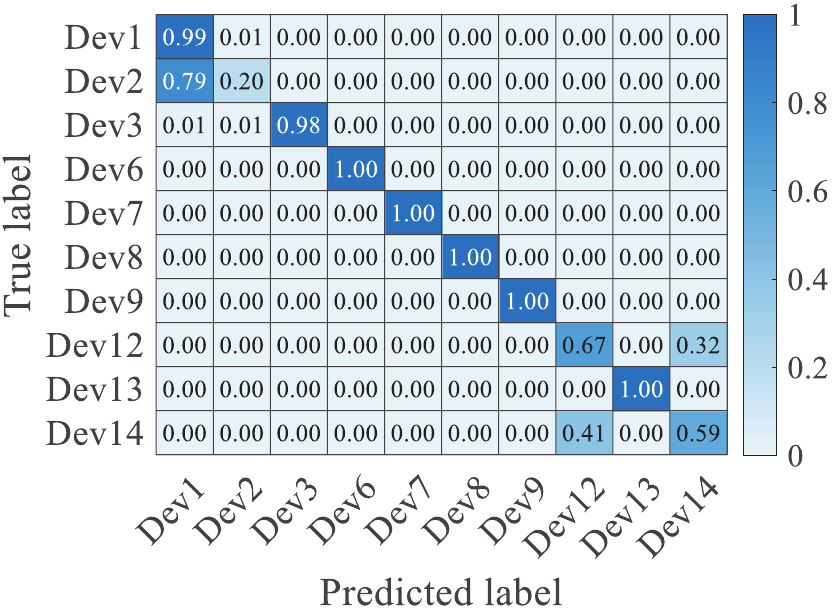}
\vspace{-0.1cm}
\caption{The confusion matrix for the proposed HL method under the static LOS scenario, with training and test sets collected by Rx2 and Rx3, respectively.}
\label{fig21}
\vspace{-0.2cm}
\end{figure}

The classification accuracy in the static corridor scenario, when trained on a single receiver and tested on Rx1 and Rx3, is shown in Fig. \ref{fig19}. As illustrated in Fig. \ref{fig19}, the classification accuracy of the proposed HL method exceeds 75\% even under receiver changes. Moreover, the proposed HL method outperforms the existing DV method by 14.25\% to 54.97\% under receiver differences.

Fig. \ref{fig20} shows the RFF features of Dev13 received by different receivers using the DV method. Comparing Fig. \ref{fig20}(a) and Fig. \ref{fig20}(b) with Fig. \ref{fig7}(b) and Fig. \ref{fig7}(d), it is evident that the RFF features extracted by the DV method are less stable than those extracted by the HL method.

We combine the features by dividing the L-STF part with the L-LTF part and dividing the HT-LTF part with the L-LTF part. Then, the classification accuracy is compared with the HL method. As shown in Fig. \ref{fig16}, Fig. \ref{fig17}, and Fig. \ref{fig18}, the classification accuracies of the combined features are similar to those of the HL method. This is because the RFF features extracted by the HL method already possess sufficient representational ability to achieve high classification accuracy.

\begin{figure}[htbp]
    \centering
    \subfloat{
        \includegraphics[width=4.18cm]{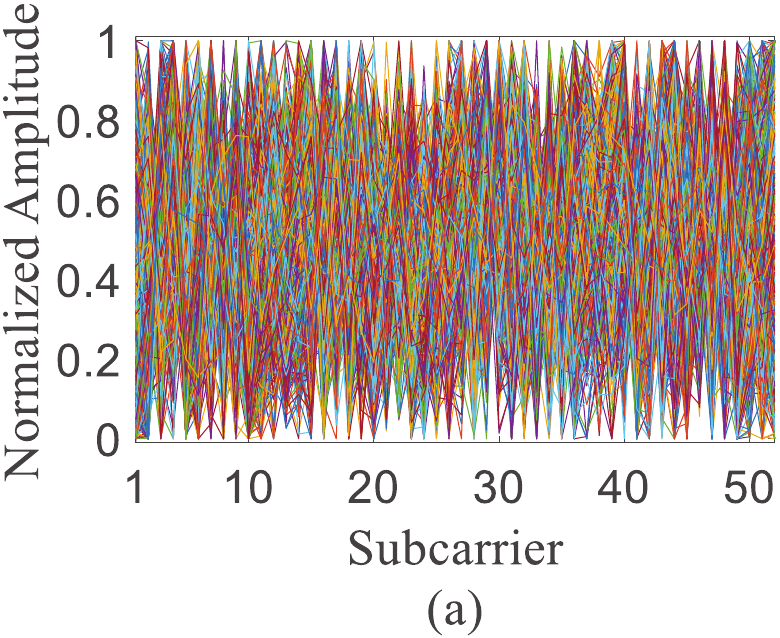}
        }\hfill
    \subfloat{
        \includegraphics[width=4.18cm]{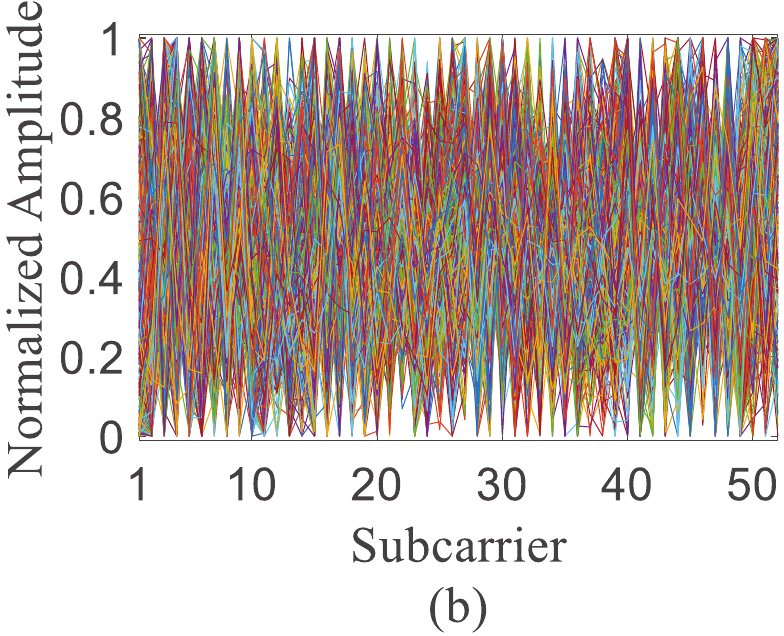}
        }
    \vspace{-0.1cm}
    \caption{RFF features of different devices extracted by Rx2 using the HL method under the static LOS scenario: (a) Dev12 and (b) Dev14.}
    \label{fig22}
    \vspace{-0.3cm}
\end{figure}

Fig. \ref{fig21} shows the confusion matrix of the proposed HL method. The training and test sets were collected by Rx2 and Rx3, respectively. As shown in Fig. \ref{fig21}, most devices achieve 100\% classification accuracy. However, Dev2 is often misclassified as Dev1 due to both devices having the same module. Similarly, Dev12 and Dev14 are frequently misclassified as each other. This is because the RFF features of these two devices in both the L-LTF and HT-LTF parts are similar, leading the HL method to extract noise-like RFF features, as shown in Fig. \ref{fig22}, which causes misclassification.
\begin{figure*}[htbp]
    \centering
    \subfloat{
        \includegraphics[width=5cm]{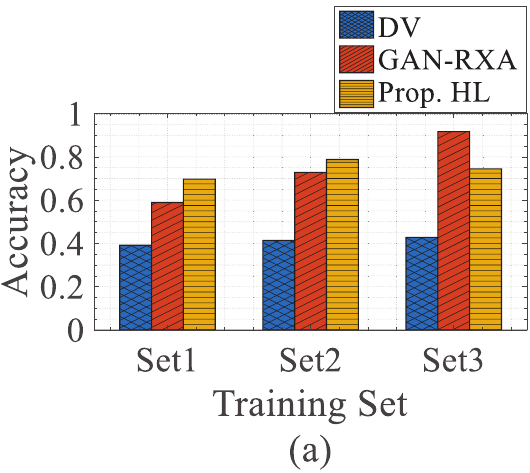}
        }\hfill
    \subfloat{
        \includegraphics[width=5cm]{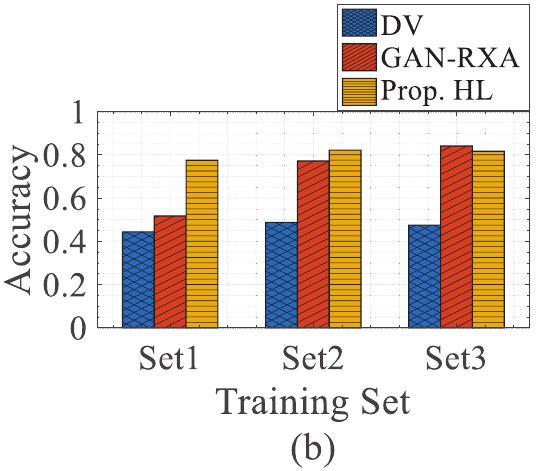}
        }\hfill
    \subfloat{
        \includegraphics[width=5cm]{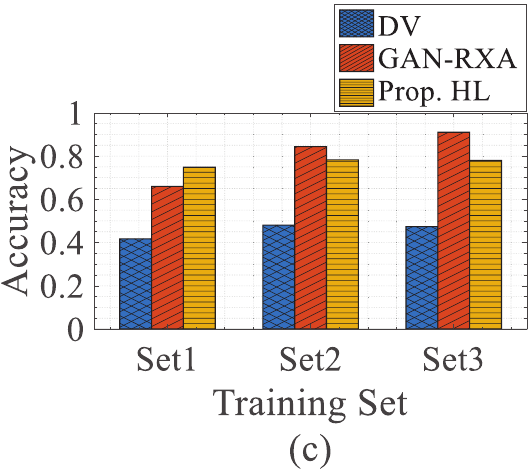}
        }
    \vspace{-0.1cm}
    \caption{Classification accuracy with training on multiple transmitters under the frequency-selective channel scenario: (a) The static LOS scenario, (b) the static NLOS scenario, and (c) the mobile scenario.}
    \label{fig23}
    \vspace{-0.3cm}
\end{figure*}
\begin{figure*}[htbp]
    \centering
    \subfloat{
        \includegraphics[width=5cm]{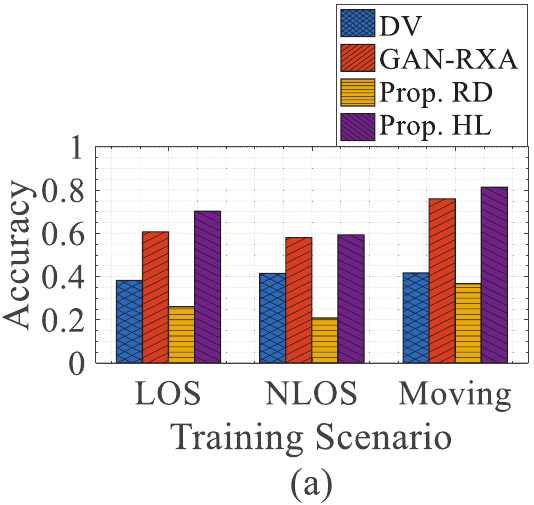}
        }\hfill
    \subfloat{
        \includegraphics[width=5cm]{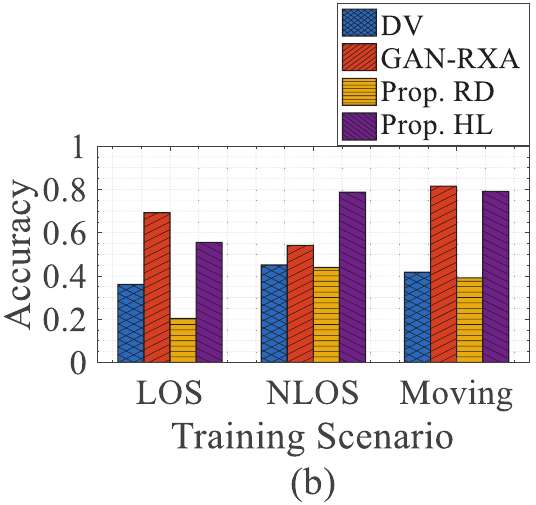}
        }\hfill
    \subfloat{
        \includegraphics[width=5cm]{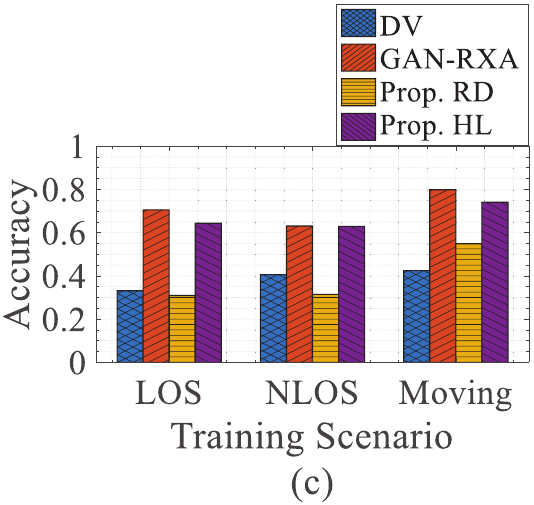}
        }
    \vspace{-0.1cm}
    \caption{Classification accuracy tested under different scenarios: (a) The static LOS scenario, (b) the static NLOS scenario, and (c) the mobile scenario.}
    \label{fig24}
    \vspace{-0.3cm}
\end{figure*}

Fig. \ref{fig23}(a), Fig. \ref{fig23}(b), and Fig. \ref{fig23}(c) show the classification accuracy of different methods in the packed indoor office when multiple receivers are used under the static LOS, static NLOS, and mobile scenarios, respectively. The test set comes from Rx1. As illustrated in Fig. \ref{fig23}, the classification accuracy of the proposed HL method is consistently higher than that of the DV method, since the HL method can extract a higher dimension of receiver-agnostic RFF features. As the number of training receivers increases, the classification accuracy of the GAN-RXA method improves and exceeds that of the HL method when Set3 is used as the training set. However, when the number of training receivers is small, e.g., two training receivers, the classification accuracy of the HL method is higher than that of the GAN-RXA method. Moreover, the training process of the neural network in the GAN-RXA method is time-consuming and prone to convergence issues, making the method less practical.

\subsubsection{Classification under Different Channel Scenarios}
Fig. \ref{fig24} shows the classification accuracies of different methods in the packed indoor office when the training and testing scenarios differ, with the training set from Set1 and the test set from Rx1.
\begin{figure}[htbp]
    \centering
    \subfloat{
        \includegraphics[width=4.18cm]{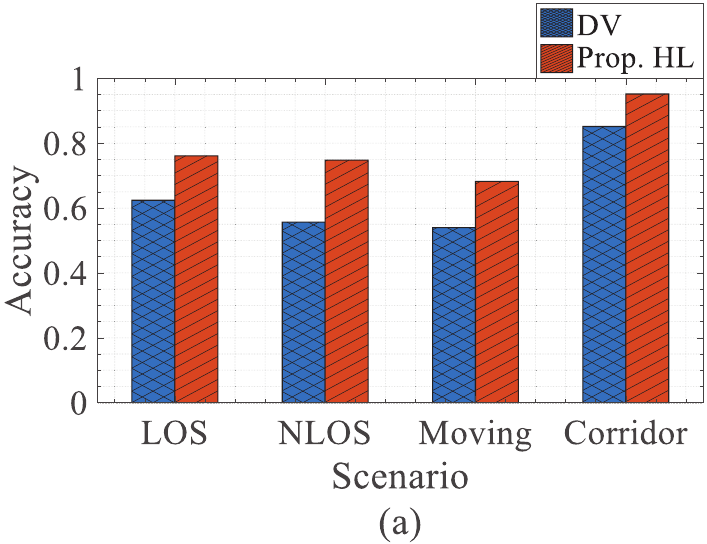}
        }\hfill
    \subfloat{
        \includegraphics[width=4.18cm]{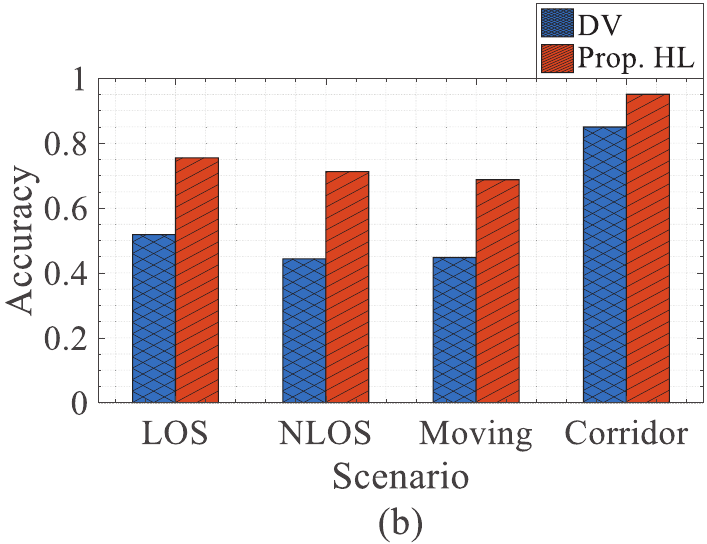}
        }
    \vspace{-0.1cm}
    \caption{Classification accuracy with training under the static corridor scenario and testing under various scenarios: (a) Under the same receiver and (b) under different receivers.}
    \label{fig25}
    \vspace{-0.3cm}
\end{figure}
As shown in Fig. \ref{fig24}, the changes in scenarios results in a decrease in the classification accuracy of the proposed HL method. This is because a small amount of channel component remains in the extracted RFF features, leading to the decrease of classification accuracy under different channel scenarios. The classification accuracy is consistently high when training in the mobile scenario, as signals collected in such scenario experience a broader range of channel conditions, which makes the trained neural network more robust to channel variations. Meanwhile, the HL method achieves the highest classification accuracy when testing on the static LOS scenario. Although the classification accuracy of the HL method is somewhat lower than that of the GAN-RXA method in certain cases, such as testing on the static NLOS and mobile scenarios, the HL method only requires one receiver for training, making it more time-efficient in signal collection and more practical for deployment. On the other hand, the RD method consistently yields lower classification accuracy, as it fails to mitigate the frequency-selective fading channel effects. In particular, as shown in Fig. \ref{fig24}(c), the proposed HL method achieves higher classification accuracy than the existing DV method when trained in the LOS or NLOS scenarios and tested in the mobile scenario. Therefore, the proposed HL method is more robust to channel variations than the existing DV method.

Classification accuracies trained under the static corridor scenario and tested under various scenarios are shown in Fig. \ref{fig25}. As illustrated in Fig. \ref{fig25}, the proposed HL method consistently achieves higher classification accuracy than the existing DV method under scenario changes. Furthermore, by comparing Fig. \ref{fig25}(a) with Fig. \ref{fig25}(b), it can be observed that the classification accuracy of the proposed method slightly decreases when different receivers are used for training and testing. In contrast, the decrease in the classification accuracy of the existing DV method is more pronounced under receiver changes. Therefore, the proposed HL method is more robust to receiver changes and spatial channel variations than the DV method.

\section{Conclusion}
\label{section:6}
In this paper, we proposed a division-based receiver-agnostic RFF extraction method in WiFi systems with only one receiver used for training while the calibration and stacking processes were not needed. In flat fading channel scenarios, a reference device based receiver-agnostic RFF identification method was designed for RFF extraction and identification. By dividing the preamble of the device to be classified and the reference device in the frequency domain, the receiver-dependent effects can be removed, and high-dimensional RFF features can be extracted. By combining the RFF features extracted from the L-STF and L-LTF parts, the proposed RD method can achieve high classification accuracy. Meanwhile, to achieve high classification accuracy under receiver differences, the device with a high $\eta _{\mathrm{LF}}$ value under the training receiver should be chosen as the reference device. In frequency-selective fading channel scenarios, a division-based receiver-agnostic HL method for RFF identification is put forward without the reference device. By dividing the HT-LTF part by the L-LTF part in the frequency domain, high dimensions of channel irrelevant and receiver-agnostic RFF features can be extracted. The simulation and experimental results demonstrated that the proposed method can extract receiver-agnostic RFF features effectively and can achieve high classification accuracy under receiver differences. As a powerful time-series classifier, InceptionTime is well suited for RFF identification using features extracted from CSI.
\bibliographystyle{IEEEtran}
\bibliography{references}

\end{document}